# Resilient Energy Efficient Healthcare Monitoring Infrastructure with Server and Network Protection

Ida Syafiza M. Isa, Taisir E.H. El-Gorashi, Mohamed O.I. Musa, and J.M.H. Elmirghani

**Abstract— In this paper, a 1 +1 server protection scheme is considered where two servers, a primary and a secondary processing server are used to serve ECG monitoring applications concurrently. The infrastructure is designed to be resilient against server failure under two scenarios related to the geographic location of primary and secondary servers and resilient against both server and network failures. A Mixed Integer Linear Programming (MILP) model is used to optimise the number and locations of both primary and secondary processing servers so that the energy consumption of the networking equipment and processing are minimised. The results show that considering a scenario for server protection without geographical constraints compared to the non-resilient scenario has resulted in both network and processing energy penalty as the traffic is doubled. The results also reveal that increasing the level of resilience to consider geographical constraints compared to case without geographical constraints resulted in higher network energy penalty when the demand is low as more nodes are utilised to place the processing servers under the geographic constraints. Also, increasing the level of resilience to consider network protection with link and node disjoint selection has resulted in a low network energy penalty at high demands due to the activation of a large part of the network in any case due to the demands. However, the results show that the network energy penalty is reduced with the increasing number of processing servers at each candidate node. Meanwhile, the same energy for processing is consumed regardless of the increasing level of resilience as the same number of processing servers are utilised. A heuristic is developed for each resilience scenario for real-time implementation where the results show that the performance of the heuristic is approaching the results of the MILP model.**

## I. INTRODUCTION

Cloud computing technologies provide services for computation and storing at anytime and anywhere. However, offloading massive amounts of data generated by end devices to the cloud for computation requests increases the congestion in the network and also increases the energy consumption of both the networking and processing equipment. Many research efforts focused on developing energy efficient architectures for cloud data centres and core networks under increasing applications' traffic [1]–[16]. To improve network energy efficiency different techniques and technologies are considered such as virtualization [6], [11], [17] , network architecture design and optimisation [14], [18], [19], [20], [21] optimising content distribution [12], [13], [22], progressive big data processing [3], [5], [7], [23], network coding [4], [16] and using renewable energy [15]. Also, fog network that integrates distributed edge servers for decentralized architecture are proposed to reduce the burden on central data centers [15], [24], [25]. In our previous work in [26], we have shown that there is 68% total energy saving when using fog computing to serve Electrocardiogram (ECG) monitoring applications to save the heart patients within the time constraint imposed by the American Heart Association (AHA) compared to the traditional cloud computing approach.

Many approaches have been introduced to improve service resilience at the cloud networking infrastructure as surveyed in [27] which range from designing and operating the facilities, servers, networks, to their integration and virtualisation. In [28], the concept of virtualization is used to allows the sharing of backup servers in geo-distributed data centers which improved the utilisation of backup servers by 40%. However, the proposed shared protection scheme requires high reserved bandwidth and can increase the latency of the secondary path between the primary and backup servers. Meanwhile, the work in [29] studied the impact of the relocation of the primary and backup servers on the total cost of both servers and network capacity. The study revealed that considering protection against single link failures with relocation reduces the cost of both servers and link capacity. Furthermore, the study showed that the benefits of relocation are more noticeable for sparser topologies. The consideration of fog computing to perform local processing at the edge network, also improves services resilience. This has been studied in [30], whereby based on their simulation, fog computing can improve network resilience by offering local processing at the network edge which also provides better response time compared to a cloud-only architecture especially for an interactive request. However, to the best of our knowledge, no work has focused on reducing the energy consumption of both networking equipment and processing while improving the service resilience while considering the server and network protection at the fog networking infrastructure level.

In our previous work in [31], we have proposed a resilient fog computing infrastructure for health monitoring application considering a 1+1 protection scheme where two servers, a primary server and a secondary server are used to serve the ECG monitoring applications concurrently in West Leeds, United Kingdom. The patients will send the necessary data to the primary and secondary processing servers for processing, analysis and decision making. A Mixed Integer Linear Programming (MILP) model was used to optimise the number and location of the processing servers to reduce the energy consumption of both networking and processing equipment. We present preliminary results to demonstrate that only network energy consumption is affected when increasing the level of resilience while considering geographical constraints compared to without geographical constraints consideration for server protections. The current paper makes several new contributions beyond those presented in [31]:

(i) It considers a wide range for the number of processing servers per candidate fog node (i.e. from 1 up to 8 processing server per candidate note) to study the energy consumed by the networking and processing equipment under both without and with geographical constraints.

(ii) Compared to [31], it provides the MILP model formulation for both scenarios without and with geographical constraints for server protection.

(iii) It provides the results of network and processing energy penalty of a resilient scenario, without geographical constraint compared to the non-resilient scenario.

(iv) It presents our new Energy optimised resilient infrastructure fog computing without geographical constraints (EORIWG) heuristic and Energy optimised resilient infrastructure fog computing with geographical constraints (EORIG) heuristic and their performance gaps compared to the MILP model results in terms of total energy consumption of both networking equipment and processing.

(v) It further extends our infrastructure to be resilient against both server and network failures. We consider the protection of servers with geographical constraints and the protection of the network with disjoint links and nodes selection, offering a scenario with higher levels of resilience. In this scenario, the primary and secondary processing servers are not allowed to be placed at the same node, and the links and nodes used to transmit the data to and from primary and secondary processing servers are disjoint, as node and link failures in the network are not improbable. We consider the disjoint links and nodes only at the access layer, as the processing servers can only be placed at the access layer. A MILP model is used to optimise the number and locations of both primary and secondary processing servers so that the energy consumption of the networking equipment and processing are minimised. Also, an Energy optimised resilient infrastructure fog computing with geographical constraints and link and node disjoint (EORIGN) heuristic is developed for real-time implementation.

The remaining of the paper is organized as follows: Section II elaborates the proposed resilient fog computing architecture for the health monitoring application. Section III depicts the mathematical modelling of the proposed approach that was developed by using MILP model considering GPON network at the access layer. Next, the parameters selection that has been considered in this work are elaborated in Section IV. The evaluation of the performance displayed by the developed approaches for health monitoring application is presented in Section V. The development of the heuristic for each considered protection scenarios is explained in Section VI while the performance evaluations of the heuristics are presented in Section VII. Finally, this paper is concluded in Section VIII.

## II. Resilient Fog computing architecture for health monitoring applications with GPON access network

The architecture for the resilient fog computing infrastructure for healthcare applications over the Gigabit Passive Optical Networks (GPON) network consists of four layers, as shown in Figure 1 [31]. The first layer (layer 1) presents the Internet of Things (IoT) layer. The second layer (layer 2) is the access layer. The third layer (layer 3) is the metro layer while the fourth layer (layer 4) is the core layer. The details of each layer are as explained in [31].

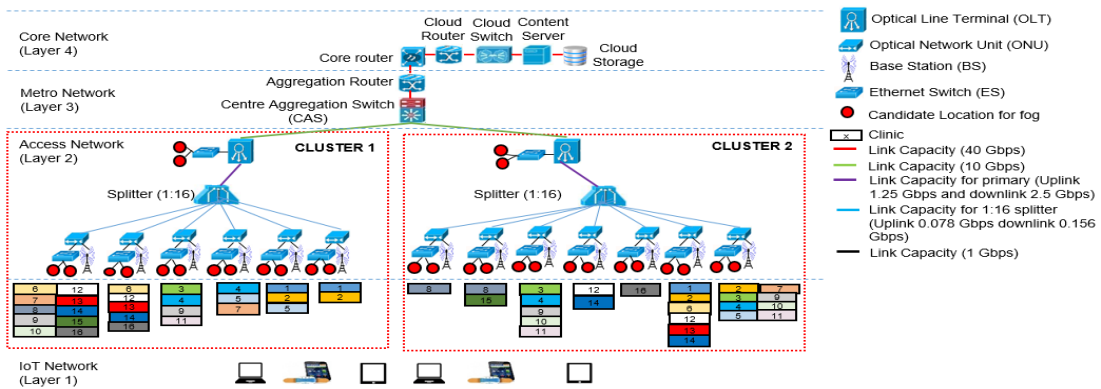

Figure 1: The resilient fog computing infrastructure under GPON access network for health monitoring applications [31]

## III. MILP MODEL FOR ENERGY-EFFICIENT AND RESILIENT INFRASTRUCTURE FOG COMPUTING HEALTH MONITORING APPLICATIONS

In this section, the mathematical model for the resilience scenarios related to the geographic location for server protection with the objective function, to minimise the total energy consumption of both networking equipment and processing are provided. Also, the mathematical model for both server and network protection that considers geographic location and link and node disjoint, respectively with the same objective function is also provided. Note that, the energy consumption of networking equipment includes the energy consumed by all networking devices at all layers while the processing energy consumption refers to the energy consumed by the primary and secondary processing servers.

### A. Protection for server without geographical constraints

To model the energy consumption minimised approach considering server protection without geographical constraints, the sets, parameters, variables and objective function are introduced as below:

Table 1: The sets, parameters and variables used in MILP

| Sets | |
|---|---|
| $CL$ | Set of clinics |
| $BS$ | Set of BSs |
| $ONU$ | Set of ONUs |
| $OLT$ | Set of OLTs |
| $CAS$ | Set of centre aggregation switches |
| $AR$ | Set of aggregation routers |
| $CR$ | Set of core routers |
| $CLR$ | Set of cloud routers |
| $CLS$ | Set of cloud switches |
| $CS$ | Set of content servers |
| $CST$ | Cloud storage |
| $N_m$ | Set of neighbouring nodes of node $m$ in the network |
| $N$ | Set of nodes ($N \in CL \cup BS \cup ONU \cup OLT \cup CAS \cup AR \cup CR \cup CLR \cup CLS \cup CS \cup CST$) |
| $FN$ | Set of candidate locations to deploy PS (fog) ($FN \in ONU \cup OLT$) |
| **Parameters** | |
| $s$ and $d$ | Denote source node $s$ and destination node $d$ of traffic between a node pair |
| $i$ and $j$ | Denote end nodes of a physical link in the network, $i, j \in N$ |
| $Pt_s$ | Number of patients in clinic $s$ |
| $IBS$ | Idle power consumption of a base station (W) |
| $PBS$ | Power per physical resource block (PRB) of a base station (W/PRB) |
| $R$ | Maximum number of PRBs in a base station dedicated for healthcare applications |
| $PONU$ | Maximum power consumption of an ONU (W) |
| $IONU$ | Idle power consumption of an ONU (W) |
| $CONU$ | Maximum capacity of an ONU (bps) |
| $PES$ | Maximum power consumption of an Ethernet switch (W) |
| $IES$ | Idle power consumption of an Ethernet switch (W) |
| $CES$ | Maximum capacity of an Ethernet switch (bps) |
| $POLT$ | Maximum power consumption of an OLT (W) |
| $IOLT$ | Idle power consumption of an OLT (W) |
| $COLT$ | Maximum capacity of an OLT (bps) |
| $PCAS$ | Maximum power consumption of a centre aggregation switch (W) |
| $ICAS$ | Idle power consumption of a centre aggregation switch (W) |
| $CCAS$ | Maximum capacity of a centre aggregation switch (bps) |
| $PAR$ | Maximum power consumption of an aggregation router (W) |
| $IAR$ | Idle power consumption of an aggregation router (W) |
| $CAR$ | Maximum capacity of an aggregation router (bps) |
| $PCR$ | Maximum power consumption of a core router (W) |
| $ICR$ | Idle power consumption of a core router (W) |
| $CCR$ | Maximum capacity of a core router (W) |
| $PCLR$ | Maximum power consumption of a cloud router (W) |
| $ICLR$ | Idle power consumption of a cloud router (W) |
| $CCLR$ | Maximum capacity of a cloud router (W) |
| $PCLS$ | Maximum power consumption of a cloud switch (W) |
| $ICLS$ | Idle power consumption of a cloud switch (W) |
| $CCLS$ | Maximum capacity of a cloud switch (bps) |
| $PCS$ | Maximum power consumption of a content server (W) |
| $ICS$ | Idle power consumption of a content server (W) |
| $CCS$ | Maximum capacity of a content server (bps) |
| $PCST$ | Maximum power consumption of a cloud storage (W) |
| $ICST$ | Idle power consumption of a cloud storage (W) |
| $CCST$ | Maximum capacity of a cloud storage (bit) |
| $PPS$ | Maximum power consumption of a processing server (W) |
| $IPS$ | Idle power consumption of a processing server (W) |
| $\Lambda u$ | Maximum number of patients per processing server |
| $\Lambda max$ | Maximum storage capacity of processing server (bit) |
| $\delta a$ | Data rate per patient to send raw health data from clinic to processing server (bps) |
| $\tau a$ | Transmission time per patient to send raw health data from clinic to processing server (s) |
| $Rp$ | Physical resource block per patient to send raw health data from clinic to processing server |
| $\alpha$ | Size of analysed health data per patient (bit) |
| $\delta b$ | Data rate per patient to send analysed health data from processing server to clinic (bps) |
| $\tau b$ | Transmission time per patient to send analysed health data from processing server to clinic (s) |
| $Rf$ | Physical resource block per patient to send analysed health data from processing server to clinic |
| $\delta c$ | Data rate per patient to send analysed health data from processing server to cloud storage (bps) |
| $\tau c$ | Transmission time per patient to send analysed health data from processing server to cloud storage (s) |
| $\delta_{sd}$ | $\delta_{sd} = 1$ to send the storage traffic from processing servers located at node $s$, to the cloud storage node $d$, $s \in FN, d \in CST$ |
| $x$ | Fraction of idle power consumption of networking equipment contributed by the healthcare application under consideration |
| $\lambda_{ij}$ | The capacity of link $ij$ dedicated for the healthcare application under consideration (bps) |
| $\eta$ | Power usage effectiveness (PUE) of the access network, metro network and IP over WDM network |
| $c$ | Power usage effectiveness (PUE) of the fog (processing server) and cloud equipment |
| $M$ | A large enough number |
| **Variables** | |
| $P_{sd}$ | Raw health data traffic from source node $s$ to destination node $d$ (bps), $s \in CL, d \in FN$ |
| $P_{ij}^{sd}$ | Raw health data traffic from source node $s$ to destination node $d$ that traverses the link between nodes $i$ and $j$ (bps), $s \in CL, d \in FN, i, j \in N$ |
| $P_i$ | Total raw health data traffic that traverses node $i$ (bps), $i \in N$ |
| $F_{sd}$ | Analysed health data feedback traffic from source node $s$ to destination node $d$ (bps), $s \in FN, d \in CL$ |
| $F_{ij}^{sd}$ | Analysed health data feedback traffic from source node $s$ to destination node $d$ that traverses the link between nodes $i$ and $j$ (bps), $s \in FN, d \in CL, i, j \in N$ |
| $F_i$ | Total analysed health data feedback traffic that traverses node $i$ (bps), $i \in N$ |
| $S_{sd}$ | Analysed health data storage traffic from source node $s$ to destination node $d$ (bps), $s \in FN, d \in CST$ |
| $S_{ij}^{sd}$ | Analysed health data storage traffic from source node $s$ to destination node $d$ that traverses the link between nodes $i$ and $j$ (bps), $s \in FN, d \in CST, i, j \in N$ |
| $S_i$ | Total analysed health data storage traffic that traverses node $i$ (bps), $i \in N$ |
| $\omega a_{sd}$ | Number of patients from clinic $s$ served by primary processing servers located at node $d$, $s \in CL, d \in FN$ |
| $\omega b_{sd}$ | Number of patients from clinic $s$ served by secondary processing servers located at node $d$, $s \in CL, d \in FN$ |
| $Pp_{ij}$ | Number of patients in clinic $i$ served by BS $j$ to send raw health data traffic (integer) |
| $Pf_{ij}$ | Number of patients in clinic $i$ served by BS $j$ to receive analysed health data feedback traffic (integer) |
| $\beta a_j$ | Number of PRBs used in BS $j$ to serve raw health data traffic (integer) |
| $\beta b_i$ | Number of PRBs used in BS $i$ to serve analysed health data feedback traffic (integer) |
| $Y_d$ | $Y_d = 1$, if a processing server is placed at node $d$, otherwise $Y_d = 0$, $d \in FN$ |
| $Ya_d$ | $Ya_d = 1$, if one or more primary processing servers are located at node $d$, otherwise $Ya_d = 0$, $d \in FN$ |
| $Yb_d$ | $Yb_d = 1$, if one or more secondary processing servers are placed at node $d$, otherwise $Yb_d = 0$, $d \in FN$ |
| $z_d$ | $z_d$ is a dummy variable that takes a value of $Ya_d \oplus Yb_d$, where $\oplus$ is an XOR operation, $d \in FN$ |
| $\phi a_d$ | Number of primary processing servers placed at node $d$, $d \in FN$ |
| $\phi b_d$ | Number of secondary processing servers placed at node $d$, $d \in FN$ |
| $\tau p a_d$ | Processing and analysis time of primary processing server (seconds) at node $d$, $d \in FN$ |
| $\tau p b_d$ | Processing and analysis time of secondary processing server (seconds) at node $d$, $d \in FN$ |
| $\zeta a_j$ | $\zeta a_j = 1$, if raw health data traffic traverses node $j$, otherwise $\zeta a_j = 0$, $j \in N$ |
| $\zeta b_i$ | $\zeta b_i = 1$, if analysed health data feedback traffic traverses node $i$, otherwise $\zeta b_i = 0$, $i \in N$ |
| $\theta_i$ | $\theta_i = 1$, if analysed health data storage traffic traverses node $i$ where node $i$ is the source of a link, otherwise $\theta_i = 0$, $i \in N$ |
| $\vartheta_j$ | $\vartheta_j = 1$, if analysed health data storage traffic traverses node $j$ where $j$ is the end of a link, otherwise $\vartheta_j = 0$, $j \in N$ |
| $\zeta c_i$ | $\zeta c_i = 1$, if the analysed health data storage traffic traverses node $i$ where $\zeta c_i = \theta_i$ OR $\vartheta_i$, otherwise $\zeta c_i = 0$, $i \in N$ |
| $\nu_i$ | $\nu_i$ is a dummy variable that takes value of $\theta_i \oplus \vartheta_i$, where $\oplus$ is an XOR operation, $i \in N$ |
| $EAN$ | Energy consumption of access network |
| $ETBS$ | Total energy consumption of base stations |
| $EBSP$ | Energy consumption of base stations required to relay raw health data traffic |
| $EBSF$ | Energy consumption of base stations required to relay analysed health data feedback traffic |
| $ETONU$ | Total energy consumption of ONUs |
| $EONUP$ | Energy consumption of ONUs required to relay raw health data traffic |
| $EONUF$ | Energy consumption of ONUs required to relay analysed health data feedback traffic |
| $EONUS$ | Energy consumption of ONUs required to relay analysed health data storage traffic |
| $ETES$ | Total energy consumption of Ethernet switches |
| $EESP$ | Energy consumption of Ethernet switches required to relay raw health data traffic |
| $EESF$ | Energy consumption of Ethernet switches required to relay analysed health data feedback traffic |
| $ETOLT$ | Total energy consumption of OLTs |
| $EOLTP$ | Energy consumption of OLTs required to relay raw health data traffic |
| $EOLTF$ | Energy consumption of OLTs required to relay analysed health data feedback traffic |
| $EOLTS$ | Energy consumption of OLTs required to relay analysed health data storage traffic |
| $EMN$ | Energy consumption of metro network |
| $ECASS$ | Energy consumption of centre aggregation switches required to relay analysed health data storage traffic |
| $EARS$ | Energy consumption of aggregation routers required to relay analysed health data storage traffic |
| $ECN$ | Energy consumption of core network |
| $ECRS$ | Energy consumption of core routers required to relay analysed health data storage traffic |
| $ECL$ | Energy consumption of cloud |
| $ECLRS$ | Energy consumption of cloud routers required to relay analysed health data storage traffic |
| $ECLSS$ | Energy consumption of cloud switches required to relay analysed health data storage traffic |
| $ECSS$ | Energy consumption of content servers required to relay analysed health data storage traffic |
| $ECSTS$ | Energy consumption of cloud storage required to store the analysed health data storage traffic |
| $EFN$ | Energy consumption of fog nodes |
| $EPS$ | Energy consumption of processing servers |

We start by defining the energy consumption of the network including access, metro and core and processing servers. Note that, in this work, we considered the power consumption of all networking equipment and processing server consists of an idle part and a linear proportional part. The Ethernet switches and the processing server are dedicated for healthcare applications (i.e. unshared) while the other devices are shared by multiple applications. Therefore, we only consider a fraction of idle power contributed by the healthcare application $(x)$ for the shared devices while maximum idle power for the unshared devices. Also, note that the energy consumed by all devices is proportional to the time the devices are utilised and the load of the devices to serve the workload:

*a)* Total energy consumption of access network, $EAN$:

The energy consumption of access network, $EAN$ is composed of energy of LTE base stations, ONUs and OLTs as given in (1):

$$EAN = (ETBS + EOTNU + ETOLT)\,\eta \qquad (1)$$

where $\eta$ is the network PUE. The energy consumption of access network had been composed of energy consumed by three different tasks i.e. processing task, feedback task and storage task that occurs at different times. In the processing task, the raw health data is sent from clinic to PS at fog. In the feedback and storage task, analysed health data is sent from the PS to the clinics and to the cloud storage, respectively. The energy consumption of LTE base stations (BSs), $ETBS$ is calculated as in Equation (2) below:

$$ETBS = EBSP + EBSF \qquad (2)$$

$$EBSP = \sum_{i \in BS}(IBS\,x\,\zeta a_i + PBS\,\beta a_i)\,\tau a \qquad (3)$$

$$EBSF = \sum_{i \in BS}(IBS\,x\,\zeta b_i + PBS\,\beta b_i)\,\tau b \qquad (4)$$

The energy consumed by LTE base stations had been based on the energy consumed to relay the raw health data traffic and the energy consumed to transmit the analysed health data feedback traffic. The energy is calculated on the number of PRBs and the time the BS is used to send the traffic as shown in equation (3)-(4).

The energy consumption of ONUs is given as follow:

$$ETONU = EONUP + EONUF + EONUS \qquad (5)$$

where

$$EONUP = \sum_{i \in ONU}\left(IONU\,x\,\zeta a_i + P_i\,\frac{(PONU - IONU)}{CONU}\right)\tau a \qquad (6)$$

$$EONUF = \sum_{i \in EONU} \left( IONU \times \zeta b_i + F_i \frac{(PONU - IONU)}{CONU} \right) \tau b \qquad (7)$$

$$EONUS = \sum_{i \in EONU} \left( IONU \times \zeta c_i + S_i \frac{(PONU - IONU)}{CONU} \right) \tau c \qquad (8)$$

The energy consumption of ONUs is calculated based on the energy consumed to relay the raw health data traffic, analysed health data feedback traffic, and analysed health data storage traffic as shown in equation (6)-(8), respectively.

The energy consumed by the OLT is given as below:

$$ETOLT = EOLTP + EOLTF + EOLTS \qquad (9)$$

where

$$EOLTP = \sum_{i \in EOLT} \left( IOLT \times \zeta a_i + P_i \frac{(POLT - IOLT)}{COLT} \right) \tau a \qquad (10)$$

$$EOLTF = \sum_{i \in EOLT} \left( IOLT \times \zeta b_i + F_i \frac{(POLT - IOLT)}{COLT} \right) \tau b \qquad (11)$$

$$EOLTS = \sum_{i \in EOLT} \left( IOLT \times \zeta c_i + S_i \frac{(POLT - IOLT)}{COLT} \right) \tau c \qquad (12)$$

The energy consumed by the OLT is calculated based on the energy consumed to relay the raw health data traffic, analysed health data feedback traffic, and analysed health data storage traffic as shown in equation (10)-(12), respectively

The energy consumption of the metro network, $EMN$ is composed of energy consumption of centre aggregation switches and aggregation routers. Note that the aggregation routers are only used to relay the analysed health data storage traffic as the candidate locations of the processing server are at the access layer Hence, the raw health data traffic and analysed health data feedback traffic does not traverse the aggregation routers. Meanwhile, the centre aggregation switches are used to relay the raw health data traffic, analysed health data feedback traffic and analysed health data storage traffic between different clusters. Therefore, the energy consumption of metro network is given as follow:

$$EMN = (ECASP + ECASF + ECASS + EARS) \eta \qquad (13)$$

where

$$ECASP = \sum_{i \in ECAS} \left( ICAS \times \zeta a_i + P_i \frac{(PCAS - ICAS)}{CCAS} \right) \tau a \qquad (14)$$

$$ECASF = \sum_{i \in ECAS} \left( ICAS \times \zeta b_i + F_i \frac{(PCAS - ICAS)}{CCAS} \right) \tau b \qquad (15)$$

$$ECASS = \sum_{i \in ECAS} \left( ICAS \times \zeta c_i + S_i \frac{(PCAS - ICAS)}{CCAS} \right) \tau c \qquad (16)$$

$$EARS = \sum_{i \in AR} \left( IAR \times \zeta c_i + S_i \frac{(PAR - IAR)}{CAR} \right) \tau c \qquad (17)$$

*b) Total energy consumption of core network, ECN*

The energy consumption of core network, $ECN$ is composed of energy consumption of core routers to relay the analysed health data storage traffic as given below:

$$ECN = ECRS \, \eta \qquad (18)$$

where

$$ECRS = \sum_{i \in ECR} \left( ICR \times \zeta c_i + S_i \frac{(PCR - ICR)}{CCR} \right) \tau c \qquad (19)$$

*c) Energy consumption of cloud, ECL*

The energy consumption of cloud, $ECL$, is composed of energy of cloud routers, cloud switches, content servers and cloud storage. Note that the cloud storage is used to perform the storage task while other devices are only used to relay the analysed health data storage traffic. The energy consumption of the cloud is given in equation (20):

$$ECL = (ECLRS + ECLSS + ECSS + ECSTS) \, c \qquad (20)$$

where

$$ECLRS = \sum_{i \in ECLR} \left( ICLR \times \zeta c_i + S_i \frac{(PCLR - ICLR)}{CCLR} \right) \tau c \qquad (21)$$

$$ECLSS = 2 \sum_{i \in ECLS} \left( ICLS \times \zeta c_i + S_i \frac{(PCLS - ICLS)}{CCLS} \right) \tau c \qquad (22)$$

$$ECSS = \sum_{i \in ECS} \left( ICS \times \zeta c_i + S_i \frac{(PCS - ICS)}{CCS} \right) \tau c \qquad (23)$$

$$ECSTS = 2 \sum_{i \in ECST} \left( ICST \times \zeta c_i + \frac{S_i}{2} \, \tau c \, \frac{(PCST - ICST)}{CCST} \right) \tau c \qquad (24)$$

Note that the energy consumption of the cloud switches and the cloud storage is multiplied by '2' for redundancy purposes [22]. Also note that, for cloud storage, the size of the analysed health data, $S_i$ is divided by '2' as only one analysed health data from both primary and secondary processing servers are stored.

*d) Energy consumption of fog nodes, EFN:*

The energy consumed by the fog, $EFN$, reflects the energy consumed by primary and secondary processing servers, $EPS$, and the energy consumed by the Ethernet switches, $ETES$ as given below:

$$EFN = EPS \, c + ETES \, \eta \qquad (25)$$

where

$$EPS = \sum_{d \in EFN} (IPS \, (\phi a_d + \phi b_d) \, (\tau a + \tau b + \tau c) + PPS \, (\tau p a_d + \tau p b_d) \qquad (26)$$

$$ETES = EESP + EESF + EESS \qquad (27)$$

$$EESP = \sum_{i \in EFN} \left( IES \times Y_i + P_i \frac{(PES - IES)}{CES} \right) \tau a \qquad (28)$$

$$EESF = \sum_{i \in EFN} \left( IES \times Y_i + F_i \frac{(PES - IES)}{CES} \right) \tau b \qquad (29)$$

$$EESS = \sum_{i \in EFN} \left( IES \times Y_i + S_i \frac{(PES - IES)}{CES} \right) \tau c \qquad (30)$$

The idle power of both primary and secondary processing servers is calculated by considering the following: the time to receive raw health data from clinic, $\tau a$, the time to transmit the analysed health data to clinics, $\tau b$, as well as the time to transmit the analysed health data to cloud storage, $\tau c$. Note that, we assume the processing server works at its full utilisation. Therefore, the proportional energy consumption of the primary and secondary processing servers is determined considering the time to perform the processing and analysis which is $\tau p a_d$ and $\tau p b_d$, respectively. Meanwhile, the energy consumption of the Ethernet switches is calculated based on the energy consumed to serve the raw health data traffic, analysed health data feedback traffic, and analysed health data storage traffic. Note that the energy of the Ethernet switches is consumed if the utilised processing servers are connected to it ($Y_i = 1$) and more than one processing servers are allowed to be served at each candidate fog node ($\phi a_i$ and $\phi b_i$ is variable).

The model is defined as follows:
Objective:
Minimise the total energy consumption of access network, $EAN$, metro network, $EMN$, core network, $ECN$, cloud, $ECL$ and fog nodes, $EFN$, given as:

$$EAN + EMN + ECN + ECL + EFN \tag{31}$$

Subject to:

*1)* Association of patients to a processing server.

$$\omega a_{sd} \le Pt_s \, Ya_d \quad ; \, \forall s \in CL, \forall d \in FN \tag{32}$$

$$\omega b_{sd} \le Pt_s \, Yb_d \quad ; \, \forall s \in CL, \forall d \in FN \tag{33}$$

Constraints (32)-(33) are used to allocated patients from clinic $s$, to be served by the primary and secondary processing servers at fog located at node $d$, respectively. Note that, if a patient is allocated to a candidate location, this location should have fog.

$$\sum_{d \in FN} \omega a_{sd} = Pt_s \quad ; \, \forall s \in CL \tag{34}$$

$$\sum_{d \in FN} \omega b_{sd} = Pt_s \quad ; \, \forall s \in CL \tag{35}$$

Constraints (34)-(35) are to ensure that all patients at clinic $s$, are assigned to the primary and secondary processing servers located at any node $d$, respectively.

*2)* Traffic from clinics to processing server.

$$P_{sd} = (\omega a_{sd} + \omega b_{sd}) \, \delta a \quad ; \, s \in CL, d \in FN \tag{36}$$

Constraint (36) calculates the raw health data traffic from clinic $s$, to the primary and secondary processing server located at node $d$. This is based on the association of patients from the clinic to the primary processing server, $\omega a_{sd}$, the association of patients from clinic to secondary processing server, $\omega b_{sd}$ as well as the data rate provisioned for each patient, $\delta a$, to perform the transmission.

*3)* Traffic from processing server to clinics.

$$F_{sd} = (\omega a_{ds} + \omega b_{ds}) \, \delta b \quad ; \, \forall s \in FN, d \in CL \tag{37}$$

Constraint (37) calculates the analysed health data feedback traffic from the primary and secondary processing servers located at node $s$, to clinic $d$. In fact, this is based on the total number of patients in the clinic, served by the primary processing servers, $\omega a_{ds}$, the total number of patients in the clinic served by the secondary processing servers, $\omega b_{ds}$, and the data rate provisioned for each patient, $\delta b$, to perform the transmission.

*4)* Traffic from fog to cloud storage.

$$S_{sd} = \sum_{i \in CL} (\omega a_{is} + \omega b_{is}) \, \delta c \, \delta_{sd} \quad ; \, \forall s \in FN, d \in CST \tag{38}$$

Constraint (38) calculates the analysed health data storage traffic from primary and secondary processing servers located at node $s$, to cloud storage $d$. Note that in this work we only utilise one cloud storage, hence, $\delta_{sd} = 1$. In fact, this is based on the total number of patients in the clinic served by primary processing servers, $\omega a_{is}$, the total number of patients in the clinic served by secondary processing servers, $\omega b_{is}$, and the data rate provisioned for each patient, $\delta c$, to perform the transmission.

*5)* Flow conservation in the network.

$$\sum_{j \in Nm[i]: i \ne j} P_{ij}^{sd} - \sum_{j \in Nm[i]: i \ne j} P_{ji}^{sd} = \begin{cases} P_{sd} \, if \, i = s \\ -P_{sd} \, if \, i = d \\ 0 \, otherwise \end{cases} \tag{39}$$

$$s \in CL, d \in FN, i \in N$$

$$\sum_{j \in Nm[i]: i \ne j} F_{ij}^{sd} - \sum_{j \in Nm[i]: i \ne j} F_{ji}^{sd} = \begin{cases} F_{sd} \, if \, i = s \\ -F_{sd} \, if \, i = d \\ 0 \, otherwise \end{cases} \tag{40}$$

$$s \in FN, d \in CL, i \in N$$

$$\sum_{j \in Nm[i]: i \ne j} S_{ij}^{sd} - \sum_{j \in Nm[i]: i \ne j} S_{ji}^{sd} = \begin{cases} S_{sd} \, if \, i = s \\ -S_{sd} \, if \, i = d \\ 0 \, otherwise \end{cases} \tag{41}$$

$$s \in FN, d \in CST, i \in N$$

Constraints (39)-(41) ensures that the total incoming traffic is equivalent to the total outgoing traffic for all nodes in the network, except for source and destination nodes for processing, feedback and storage for tasks, respectively.

*6)* Total traffic traversing node.

$$P_i = \left( \sum_{s \in CL} \sum_{d \in FN: s \ne d} \sum_{j \in Nm[i]: i \ne j} P_{ji}^{sd} \right) \quad ; \, \forall i \in N \tag{42}$$

$$F_i = \left( \sum_{s \in FN} \sum_{d \in CL: s \ne d} \sum_{j \in Nm[i]: i \ne j} F_{ij}^{sd} \right) \quad ; \, \forall i \in N \tag{43}$$

$$S_i = \left( \sum_{s \in FN} \sum_{d \in CST: s \ne d} \sum_{j \in Nm[i]: i \ne j} S_{ji}^{sd} + \sum_{d \in CST: i \ne d} S_{id} \right) \quad ; \, \forall i \in N \tag{44}$$

Equations (42)-(44) calculate the total raw health data traffic, analysed health data feedback traffic, and analysed health data storage traffic that traverse node $i$, respectively.

*7)* Link capacity constraint.

$$\sum_{s \in CL} \sum_{d \in FN} P_{ij}^{sd} \le \lambda_{ij} \quad ; \, \forall i \in N, \forall j \in Nm[i]: i \ne j \tag{45}$$

$$\sum_{s \in FN} \sum_{d \in CL} F_{ij}^{sd} \le \lambda_{ij} \quad ; \, \forall i \in N, \forall j \in Nm[i]: i \ne j \tag{46}$$

$$\sum_{s \in FN} \sum_{d \in CST} S_{ij}^{sd} \le \lambda_{ij} \quad ; \, \forall i \in N, \forall j \in Nm[i]: i \ne j \tag{47}$$

Constraints (45)-(47) ensure that the capacity of physical links used to send the total raw health data from all clinics $s$ to primary and secondary processing servers at node $d$ for processing task, the total analysed health data from all processing servers at node $s$ to the clinic $d$ for feedback task, and the total analysed health data from all primary and secondary processing servers at node $s$ to the cloud storage $d$ for storage task, respectively, does not exceed the maximum capacity of the links. Note that, as mentioned above, the three tasks occur at different times.

### 8) Nodes used to connect the servers.

$$Ya_d + Yb_d = 2Y_d - z_d \quad ; \forall d \in FN \tag{48}$$

$$\phi a_d + \phi b_d \leq N \quad ; \forall d \in FN \tag{49}$$

Constraint (48), is to determine the nodes that are used to place the processing servers where $Y_d = 1$ if at least any of $Ya_d$ and $Yb_d$ are equal to 1 ($Ya_d + Yb_d$), otherwise zero. This is achieved by, introducing a binary variable $z_d$ which is only equal to 1 if $Ya_d$ and $Yb_d$ are exclusively equal to 1 ($Ya_d \oplus Yb_d$), otherwise, it is zero. Constraint (49) is to ensure that the number of processing servers at node $d$ does not exceed the maximum number of processing servers allowed at each candidate node $N$.

### 9) Node used to transmit the raw health data traffic from clinic to processing server.

$$\sum_{s \in CL} \sum_{d \in FN} \sum_{i \in N:i \neq j} P_{ij}^{sd} \geq \zeta a_j \quad ; \forall j \in N \tag{49}$$

$$\sum_{s \in CL} \sum_{d \in FN} \sum_{i \in N:i \neq j} P_{ij}^{sd} \leq M \zeta a_j \quad ; \forall j \in N \tag{50}$$

Constraints (49) and (50) ensure that $\zeta a_i = 1$ if the raw health data traffic traverses at nodes $i$ to send the data from clinic $s$ to the processing server at node $d$, otherwise it is zero.

### 10) Node used to transmit the analysed health data feedback traffic from processing server to clinic

$$\sum_{s \in FN} \sum_{d \in CL} \sum_{j \in Nm[i]:i \neq j} F_{ij}^{sd} \geq \zeta b_i \quad ; \forall i \in N \tag{51}$$

$$\sum_{s \in FN} \sum_{d \in CL} \sum_{j \in Nm[i]:i \neq j} F_{ij}^{sd} \leq M \zeta b_i \quad ; \forall i \in N \tag{52}$$

Constraints (51) and (52) ensure $\zeta b_i = 1$ if the analysed health data feedback traffic traverses node $i$ to send the analysed health data from processing servers at node $s$ to clinics $d$, otherwise it is zero.

### 11) Node used to transmit the analysed health data storage traffic from processing server to cloud storage.

$$\sum_{s \in FN} \sum_{d \in CST} \sum_{j \in Nm[i]:i \neq j} S_{ij}^{sd} \geq \theta_i \quad ; \forall i \in N \tag{53}$$

$$\sum_{s \in FN} \sum_{d \in CST} \sum_{j \in Nm[i]:i \neq j} S_{ij}^{sd} \leq M \theta_i \quad ; \forall i \in N \tag{54}$$

$$\sum_{s \in FN} \sum_{d \in CST} \sum_{j \in Nm[i]:i \neq j} S_{ij}^{sd} \geq \vartheta_j \quad ; \forall j \in N \tag{55}$$

$$\sum_{s \in FN} \sum_{d \in CST} \sum_{i \in Nm[j]:i \neq j} S_{ij}^{sd} \leq M \vartheta_j \quad ; \forall j \in N \tag{56}$$

$$\theta_i + \vartheta_i = 2 \zeta c_i - \nu_i \quad ; \forall i \in N \tag{57}$$

Constraints (53)-(54) ensure that $\theta = 1$ if the analysed health data storage traffic traverses node $i$ to send the analysed data from processing servers at node $s$ to cloud storage $d$, otherwise it is zero. However, this does not include the last node (i.e. cloud storage) that performs the storage task. Hence, constraints (55)-(56) are to ensure $\vartheta_j = 1$ if the traffic traverse node $j$ (including the last node) while constraint (57) is used to determine the activation of all nodes to relay and store the analysed health data storage traffic by ensuring that the $\zeta c_i = 1$ if at least any of $\theta_i$ and $\vartheta_i$ are equal to 1 ($\theta_i$ OR $\vartheta_i$), otherwise zero. We achieve this by introducing a binary variable $\nu_i$ which is only equal to 1 if $\theta_i$ and $\vartheta_i$ are exclusively equal to 1 ($\theta_i$ XOR $\vartheta_i$) otherwise, it is zero.

### 12) Number of physical resource blocks at each BS to send the raw health data traffic from clinics to the processing servers.

$$Pp_{ij} = \sum_{s \in CL} \sum_{d \in FN:s \neq d} \frac{P_{ij}^{sd}}{\delta a} \quad ; \forall i \in CL, \forall j \in BS: i \neq j \tag{58}$$

$$\sum_{j \in BS} Pp_{ij} = Pt_i \quad ; \forall i \in CL \tag{59}$$

$$\beta a_j = \sum_{i \in CL} Pp_{ij} \ Rp \quad ; \forall j \in BS \tag{60}$$

$$\beta a_j \leq R \quad ; \forall j \in BS \tag{61}$$

Constraint (58) is used to ensure that each patient in the clinic is served by a single BS to perform the processing task based on the traffic traversing the BS, $P_{ij}^{sd}$, and the size of raw health data traffic of each patient, $\delta a$, while constraint (59) is used to ensure that all patients are served by the BSs. Constraint (60) calculates the total number of PRBs used at each BS. Meanwhile, constraint (61) is used to ensure that the number of PRBs in each BS $j$ do not exceed their maximum number of PRBs, $R$, that are dedicated for healthcare applications to perform the processing task.

### 13) Number of physical resource blocks at each BS to send the analysed health data traffic from processing servers to clinics.

$$Pf_{ij} = \sum_{s \in FN} \sum_{d \in CL:s \neq d} \frac{F_{ij}^{sd}}{\delta b} \quad ; \forall i \in BS, \forall j \in CL \tag{62}$$

$$\sum_{i \in BS} Pf_{ij} = Pt_j \quad ; \forall j \in CL \tag{63}$$

$$\beta b_i = \sum_{j \in CL} Pf_{ij} \ Rf \quad ; \forall i \in BS \tag{64}$$

$$\beta b_i \leq R \quad ; \forall i \in BS \tag{65}$$

Constraint (62) ensures the analysed health data of each patient transmitted to the clinics is relayed by a single BS to perform the feedback task based on the traffic traversing the BS, $F_{ij}^{sd}$, and the size of analysed health data feedback traffic of each patient, $\delta b$, while constraint (63) ensures all patients are served by the BSs. Constraint (64) calculates the total number of PRBs used at each BS. Constraint (65) is used to ensure that the number of PRBs in each BS $i$ does not exceed its maximum number of PRBs, $R$, that are dedicated for healthcare applications to perform the feedback task.

### 14) Maximum number of patients served at each processing server.

$$\sum_{s \in CL} \omega a_{sd} \leq \Omega max \ \phi a_d \quad ; \forall d \in FN \tag{66}$$

$$\sum_{s \in CL} \omega b_{sd} \leq \Omega max \ \phi b_d \quad ; \ \forall d \in FN \qquad (67)$$

Constraints (66) and (67) ensures that the total number of patients served by each primary and secondary processing server at node $d$, respectively, does not exceed its maximum number of users, $\Omega max$. However, the model also allows more than one processing server, $\phi b_d$, to be deployed at the same node $d$ if the number of users is higher than $\Omega max$.

*15)* Processing and analysis time at each processing server.

$$\tau p a_d = \sum_{s \in CL} m \ \omega a_{sd} + \acute{c} \ \phi a_d \quad ; \forall d \in FN \qquad (68)$$

$$\tau p b_d = \sum_{s \in CL} m \ \omega b_{sd} + \acute{c} \ \phi b_d \quad ; \forall d \in FN \qquad (69)$$

Constraint (68) and constraint (69) is to calculate the processing and analysis time of the primary processing server and secondary processing server at node $d$, respectively. This is based on the total number of patients served by the processing server (i.e. $\omega a_{sd}$ for primary processing server and $\omega b_{sd}$ for secondary processing server) and number of processing servers used (i.e. $\phi a_d$ for primary processing server and $\phi b_d$ for secondary processing server), where $m$ and $\acute{c}$ are constant value.

*16)* Storage capacity constraint at each processing server.

$$\sum_{s \in CL} \omega a_{sd} \ \alpha \leq \Lambda max \ \phi a_d \quad ; \ \forall d \in FN \qquad (70)$$

$$\sum_{s \in CL} \omega b_{sd} \ \alpha \leq \Lambda max \ \phi b_d \quad ; \ \forall d \in FN \qquad (71)$$

Constraint (70) and constraint (71) are to ensure that the storage capacity of a primary processing server and secondary processing server at node $d$, do not exceed its maximum capacity, $\Lambda max$, respectively. Note, that the model also allows more than one primary processing servers, $\phi a_d$ and secondary processing servers, $\phi b_d$ to be deployed at the same fog processing unit located at node $d$, if the size of the data is higher than $\Lambda s$. Furthermore, this work omitted the capacity of cloud storage as a constraint. This is mainly because the storage capacity at the central cloud is large enough to sufficiently accommodate large amounts of data.

### B. Protection for servers with geographical constraints

This section considered server protection with geographical constraints, where the primary and secondary processing servers are not allowed to be placed at the same node. Typically, most service providers place their primary and secondary services in distant locations, to increase resilience. For example, BackupVault, which is a leading provider of online cloud backup for businesses in United Kingdom, locate their primary data centre in Slough, UK; while the second data centre for redundancy is located in Reading, UK [32]. Therefore, this work considered that the nodes serving the primary processing servers are not allowed to serve any of the secondary processing servers. The same parameters, variables, constraints and objective functions in the previous scenario are utilised. However, to ensure that the locations of both primary and secondary processing servers are different, constraint (48) is replaced with Equation (72), as shown below:

$$Ya_d + Yb_d = Y_d \ ; \forall d \in FN \qquad (72)$$

where constraint (72), ensures that either primary or secondary processing servers can be placed at one location $d$ as the maximum value for $Y_d$ is 1.

### C. Protection for servers with geographical constraints and protection of network with link and node disjoint

In this scenario, the primary and secondary processing servers are not allowed to be placed at the same node, and the links and nodes used to relay the traffic to and from both primary and secondary processing servers are disjoint. Beyond the OLT and heading to the cloud, the network is not protected, since the server that did the processing has a copy of the data to be stored and can retain it until the network beyond the OLT recovers. Note that the disjoint links and nodes are considered to be only at the access layer. The same parameters, variables, constraints and objective functions in Section III(B) are utilised and an additional set and variables, as shown in Table 2, are introduced to optimise the number and locations of the primary and secondary processing servers considering the geographical constraints and link and node disjoint resilience, so that the energy consumption of both networking equipment and processing are minimised.

Table 2: Additional set and variables in the MILP model

| Set | |
|---|---|
| $ND$ | Set of BSs, ONUs and OLTs (access layer) |
| **Variables** | |
| $Pa_{sd}$ | Raw health data traffic from clinic $s$ to primary processing servers at destination node $d$ (bps), $s \in CL, d \in FN$ |
| $Pb_{sd}$ | Raw health data traffic from source node $s$ to secondary processing servers at destination node $d$ (bps), $s \in CL, d \in FN$ |
| $Pa_{ij}^{sd}$ | Raw health data traffic from source node $s$ to primary processing servers at destination node $d$ that traverses the link between nodes $i$ and $j$ (bps), $s \in CL, d \in FN, \ i, j \in N$ |
| $Pb_{ij}^{sd}$ | Raw health data traffic from source node $s$ to secondary processing servers at destination node $d$ that traverses the link between nodes $i$ and $j$ (bps), $s \in CL, d \in FN, \ i, j \in N$ |
| $Fa_{sd}$ | Analysed health data feedback traffic from primary processing servers at source node $s$ to clinic at node $d$ (bps), $s \in FN, d \in CL$ |
| $Fb_{sd}$ | Analysed health data feedback traffic from secondary processing servers at source node $s$ to clinic at node $d$ (bps), $s \in FN, d \in CL$ |
| $Fa_{ij}^{sd}$ | Analysed health data feedback traffic from primary processing servers at source node $s$ to clinic at node $d$ that traverses the link between nodes $i$ and $j$ (bps), $s \in FN, d \in CL, i, j \in N$ |
| $Fb_{ij}^{sd}$ | Analysed health data feedback traffic from secondary processing servers at source node $s$ to clinic at node $d$ that traverses the link between nodes $i$ and $j$ (bps), $s \in FN, d \in CL, i, j \in N$ |
| $Sa_{sd}$ | Analysed health data storage traffic from primary processing servers at source node $s$ to cloud storage at node $d$ (bps), $s \in FN, d \in CST$ |
| $Sb_{sd}$ | Analysed health data storage traffic from secondary processing servers at source node $s$ to cloud storage at node $d$ (bps), $s \in FN, d \in CST$ |
| $Sa_{ij}^{sd}$ | Analysed health data storage traffic from primary processing servers at source node $s$ to cloud storage at node $d$ that traverses the link between nodes $i$ and $j$ (bps), $s \in FN, d \in CST, i, j \in N$ |
| $Sb_{ij}^{sd}$ | Analysed health data storage traffic from secondary processing servers at source node $s$ to cloud storage at node $d$ that traverses the link between nodes $i$ and $j$ (bps), $s \in FN, d \in CST, i, j \in N$ |
| $La_{ij}$ | $La_{ij} = 1$, if the incoming and/or outgoing traffic of primary processing servers traverses the link between nodes $i$ and $j$ otherwise $La_{ij} = 0$ |
| $Lb_{ij}$ | $Lb_{ij} = 1$, if the incoming and/or outgoing traffic of secondary processing servers traverses the link between nodes $i$ and $j$ otherwise $Lb_{ij} = 0$ |

| $\rho a_i$ | $\rho a_i = 1$, if the incoming and/or outgoing traffic of primary processing servers traverse node $i$, otherwise $\rho a_i = 0$ |
|---|---|
| $\rho b_i$ | $\rho b_i = 1$, if the incoming and/or outgoing traffic of secondary processing servers traverses node $i$, otherwise $\rho b_i = 0$ |

In addition to constraints presented in Section III(B), the following new constraints are considered:

1) Traffic from clinics to fog.

$$Pa_{sd} = \omega a_{sd} \; \delta a \quad ; s \in CL, d \in FN \tag{73}$$

$$Pb_{sd} = \omega b_{sd} \; \delta a \quad ; s \in CL, d \in FN \tag{74}$$

Constraints (73)-(74) calculate the raw health data traffic from clinic $s$ to the primary and secondary processing servers located at node $d$, respectively. This is based on the association of patients from clinic to processing servers (i.e. $\omega a_{sd}$ and $\omega b_{sd}$), as well as the data rate provisioned for each patient, $\delta a$, to perform the transmission.

2) Traffic from fog to clinics.

$$Fa_{sd} = \omega a_{sd} \; \delta b \quad ; s \in FN, d \in CL \tag{75}$$

$$Fb_{sd} = \omega b_{sd} \; \delta b \quad ; s \in FN, d \in CL \tag{76}$$

Constraints (75)-(76) calculate the analysed health data feedback traffic from primary and secondary processing servers located at node $s$ to the clinic $d$, respectively. This is based on the association of patients from clinic to processing servers (i.e. $\omega a_{sd}$ and $\omega b_{sd}$), as well as the data rate provisioned for each patient, $\delta b$, to perform the transmission.

3) Traffic from fog to cloud storage.

$$Sa_{sd} = \sum_{i \in CL} \omega a_{is} \; \delta c \; \delta_{sd} \quad ; s \in FN, d \in CST \tag{77}$$

$$Sb_{sd} = \sum_{i \in CL} \omega b_{is} \; \delta c \; \delta_{sd} \quad ; s \in FN, d \in CST \tag{78}$$

Constraints (77)-(78) calculate the analysed health data storage traffic from primary and secondary processing servers located at node $s$ to cloud storage, $d$ respectively. This is based on the association of patients from clinic to processing servers (i.e. $\omega a_{is}$ and $\omega b_{is}$), as well as the data rate provisioned for each patient, $\delta c$, to perform the transmission. Note that, in this work, there is only one cloud storage $d$, therefore the $\delta_{sd}$ is a parameter that is equal to 1.

4) Flow conservation in the network.

$$\sum_{j \in Nm[i]: i \neq j} Pa_{ij}^{sd} - \sum_{j \in Nm[i]: i \neq j} Pa_{ji}^{sd} = \begin{cases} Pa_{sd} \text{ if } i = s \\ -Pa_{sd} \text{ if } i = d \\ 0 \text{ otherwise} \end{cases} \tag{79}$$

$$s \in CL, d \in FN, i \in N$$

$$\sum_{j \in Nm[i]: i \neq j} Pb_{ij}^{sd} - \sum_{j \in Nm[i]: i \neq j} Pb_{ji}^{sd} = \begin{cases} Pb_{sd} \text{ if } i = s \\ -Pb_{sd} \text{ if } i = d \\ 0 \text{ otherwise} \end{cases} \tag{80}$$

$$s \in CL, d \in FN, i \in N$$

$$\sum_{j \in Nm[i]: i \neq j} Fa_{ij}^{sd} - \sum_{j \in Nm[i]: i \neq j} Fa_{ji}^{sd} = \begin{cases} Fa_{sd} \text{ if } i = s \\ -Fa_{sd} \text{ if } i = d \\ 0 \text{ otherwise} \end{cases} \tag{81}$$

$$s \in FN, d \in CL, i \in N$$

$$\sum_{j \in Nm[i]: i \neq j} Fb_{ij}^{sd} - \sum_{j \in Nm[i]: i \neq j} Fb_{ji}^{sd} = \begin{cases} Fb_{sd} \text{ if } i = s \\ -Fb_{sd} \text{ if } i = d \\ 0 \text{ otherwise} \end{cases} \tag{82}$$

$$s \in FN, d \in CL, i \in N$$

$$\sum_{j \in Nm[i]: i \neq j} Sa_{ij}^{sd} - \sum_{j \in Nm[i]: i \neq j} Sa_{ji}^{sd} = \begin{cases} Sa_{sd} \text{ if } i = s \\ -Sa_{sd} \text{ if } i = d \\ 0 \text{ otherwise} \end{cases} \tag{83}$$

$$s \in FN, d \in CST, i \in N$$

$$\sum_{j \in Nm[i]: i \neq j} Sb_{ij}^{sd} - \sum_{j \in Nm[i]: i \neq j} Sb_{ji}^{sd} = \begin{cases} Sb_{sd} \text{ if } i = s \\ -Sb_{sd} \text{ if } i = d \\ 0 \text{ otherwise} \end{cases} \tag{84}$$

$$s \in FN, d \in CST, i \in N$$

Constraints $(79) - (84)$ ensure that the total incoming traffic is equivalent to the total outgoing traffic for all nodes in the network, except for source and destination nodes for processing, feedback and storage tasks, respectively.

5) Link used to transmit raw and analysed health data traffic.

$$\sum_{s \in CL} \sum_{d \in FN} Pa_{ij}^{sd} + \sum_{s \in FN} \sum_{d \in CL} Fa_{ij}^{sd} + \sum_{s \in FN} \sum_{d \in CST} Sa_{ij}^{sd} \geq La_{ij} \tag{85}$$

$$i \in N, j \in Nm[i]$$

$$\sum_{s \in CL} \sum_{d \in FN} Pa_{ij}^{sd} + \sum_{s \in FN} \sum_{d \in CL} Fa_{ij}^{sd} + \sum_{s \in FN} \sum_{d \in CST} Sa_{ij}^{sd} \leq M \, La_{ij} \tag{86}$$

$$i \in N, j \in Nm[i]$$

$$\sum_{s \in CL} \sum_{d \in FN} Pb_{ij}^{sd} + \sum_{s \in FN} \sum_{d \in CL} Fb_{ij}^{sd} + \sum_{s \in FN} \sum_{d \in CST} Sb_{ij}^{sd} \geq Lb_{ij} \tag{87}$$

$$i \in N, j \in Nm[i]$$

$$\sum_{s \in CL} \sum_{d \in FN} Pb_{ij}^{sd} + \sum_{s \in FN} \sum_{d \in CL} Fb_{ij}^{sd} + \sum_{s \in FN} \sum_{d \in CST} Sb_{ij}^{sd} \leq M \, Lb_{ij} \tag{88}$$

$$i \in N, j \in Nm[i]$$

Constraints (85)-(86) ensure that $La_{ij} = 1$ if the incoming and/or outgoing traffic of primary processing servers traverses the link between nodes $i$ and $j$, otherwise the value is zero. Meanwhile, Constraints (87)-(88) ensure that the $Lb_{ij} = 1$, if the incoming and/or outgoing traffic of secondary processing servers traverses the link between nodes $i$ and $j$, otherwise the value is zero.

6) Disjoint links constraint.

$$La_{ij} + Lb_{ij} \leq 1 \quad ; i \in ND, j \in ND \tag{89}$$

$$La_{ij} + Lb_{ji} \leq 1 \quad ; i \in ND, j \in ND \tag{90}$$

$$La_{ji} + Lb_{ij} \leq 1 \quad ; i \in ND, j \in ND \tag{91}$$

Constraints (89)–(91) ensure that the incoming and/or outgoing traffic of the primary and secondary processing servers traverse different links.

7) Disjoint nodes constraint.

$$\sum_{j \in Nm[i]: i \neq j} La_{ij} \geq \rho a_i \quad ; i \in ND \tag{92}$$

$$\sum_{j \in Nm[i]: i \neq j} La_{ij} \leq M \rho a_i \quad ; i \in ND \tag{93}$$

$$\sum_{j \in Nm[i]: i \neq j} Lb_{ij} \geq \rho b_i \quad ; i \in ND \tag{94}$$

$$\sum_{j \in Nm[i]: i \neq j} Lb_{ij} \leq M \rho b_i \quad ; i \in ND \tag{95}$$

$$\rho a_i + \rho b_i \leq 1 \quad ; i \in ND \tag{96}$$

Constraints (92)–(93) and constraints (94)-(95) determine the nodes that are used to relay the incoming and/or outgoing traffic of the primary processing server and secondary processing servers, respectively. Meanwhile, constraint (96) ensures that the nodes used to relay the incoming and/or outgoing traffic of primary and secondary processing servers are different.

## IV. PARAMETER SELECTIONS

This section elaborates in detail the methodologies of determining the model input parameters considered in this work. The input parameters are divided into several parts, as explained in the following:

### 1) Network layout

In this study, the location of patients is considered to be at the clinic where they are registered. A total of 37 clinics were available in West Leeds in 2014/2015 [33]. However, the complexity of the MILP model grows exponentially with the number of nodes in the network. Therefore, a scenario with 16 clinics and 13 LTE base stations, is considered using the locations at West Leeds as a case study. The 13 LTE base stations are selected, based on the nearest distance between the available base stations (BSs) and the clinics. Note that the locations of clinics and BSs (i.e. latitude and longitude) refer to the actual locations found in West Leeds, which had been obtained from Google Maps based on the names of clinics listed by [33] in 2014/2015 and OFCOM UK Mobile Site finder published in May 2012 [34], respectively.

Two clusters are considered as a case study and the clinics are connected to up to two of the nearest BSs in each cluster, as shown in Figure 1. For example, clinic 13, shown in red, is connected to two base stations in cluster 1, and also a single base station in cluster 2. Note that, the BSs in each cluster are assigned as follows: We determine the BSs that has the highest distance and set them as the central point for each cluster. Then we assigned the remaining BSs to the cluster with the lowest distance. For each cluster, we choose only one OLT provided by the BT Wholesale network [35] that has the lowest total distance to the BSs in the selected cluster.

### 2) Total number of monitored patients for ECG monitoring applications

In this work, we consider patients that may experience postoperative atrial fibrillation (AF) in West Leeds, UK as the respondent which relates to the total traffic considered in the network. It has been reported by the British Heart Foundation that, the total UK population of those aged 18 years old and above who undergone the heart-related surgeries (i.e. Coronary Artery Bypass Graph (CABG) and Percutaneous Coronary Interventions (PCIs)) performed in NHS and selected private hospitals in 2014 are 113, 656 [36]. Meanwhile, the Office for National Statistic (ONS) have declared that the UK population was 63,818,387 in 2014. Therefore, 0.176% of the UK population is considered to have heart surgeries. This percentage is used to estimate the number of monitored patients registered in each considered clinic[33]. Table 3 shows the deduced total number of patients registered at each clinic who have been expected to experience postoperative AF.

Table 3: Number of monitored patients in clinics

| Clinic | Total Number of Patients |
|---|---|
| Craven Road Medical Practice | 20 |
| Leeds Student Practice | 68 |
| Hyde Park Surgery | 13 |
| Burton Croft Surgery | 15 |
| Laurel Bank Surgery | 16 |
| Kirkstall Lane Medical Centre | 11 |
| Burley Park Medical Centre | 23 |
| Thornton Medical Centre | 18 |
| Beech Tree Medical Centre | 16 |
| Hawthorn Surgery | 4 |
| Priory View Medical Centre | 20 |
| Abbey Grange Medical Centre | 9 |
| Vesper Road Surgery | 16 |
| The Highfield Medical Centre | 25 |
| Dr G Lees & Partners | 10 |
| Whitehall Surgery | 16 |

### 3) Link capacity

As reported in [37], the total IP traffic of M2M from the global IP traffic in 2021 is expected to be 5%. Meanwhile, the Cisco also reported that the total M2M connected devices and the connected health consumers of M2M connection in 2020 is 12.2 billion and 729 million, respectively [38]. This denotes that the M2M connected devices for healthcare application are approximately 6% in 2020. It is important to highlight that the link capacities at all layers (access, metro and core layers) are considered to serve traffic for all other applications. As in our previous work in [26], [31], in this work, we consider only 5% were employed to predict M2M traffic from the global traffic, while 6% signified healthcare traffic from the total M2M traffic. Therefore, 0.3% of the maximum capacities at all layers are dedicated to healthcare application.

### 4) Data rate for each patient

In this work, the 30 seconds ECG recording signal ($\Pi$) with a size of 252.8 kbits, is utilised [39],[40]. Patients send their ECG signals to the network to be processed and analysed at both primary and secondary processing servers of the fog layer. The relationship between the processing and analysis time of the signal and the number of patients to perform the processing at both processing servers utilising the Pan-Tompkins algorithm are retrieved from the experiment conducted using MATLAB with parallel processing on an Intel Core i5-4460 with 3.2 GHz CPU and 500 GByte hard drive [41]. Based on the results, the duration of processing and analysis at a given number of patients ($Pat$) is: $\tau p = 0.002\ Pat + 4.6857$.

In this work, the number of patients that can be served in a single processing server $Pat$ is limited, to investigate the distribution of primary and secondary processing servers in the network, with increasing demands. Therefore, the maximum $Pat$ that can be served at a single server is considered to be 20% of the total number of patients from the 16 clinics, which is the lowest demand evaluated in the network. Based on our experimental results, the size of the processed and analysed data $\alpha$, was found to be 256 bits. This result will be sent from the primary and secondary processing servers to the cloud for permanent storage, but only one copy will be stored. The same principle applies to the data that is sent to the clinic from both servers.

The energy consumption of networking equipment and processing is calculated, based on the timing constraints set by the American Heart Association (AHA) [25]. Hence, 4 minutes (i.e. $\tau t = 4\ minutes$) is considered as the maximum duration to save heart patients. The $4\ minutes$ include; (i) the time to record the 30-second ECG signal, $\tau m$, (ii) the time to transmit the raw ECG signals to both servers for processing task, $\tau max$, (iii) the time for processing and analysis, $\tau p$, and (iv) the time to transmit the analysed ECG data for feedback, $\tau b$. To determine the available time to transmit the raw ECG signal to the processing servers, $\tau max$, the time of both processing and analysis, $\tau p$ is calculated based on the maximum number of patients that can be served by a single processing server ($Pat$) and the time to send the analysed ECG data to the clinics for feedback, $\tau b$ while considering the 30 seconds of ECG recording, $\tau m$ from the patient for $\tau t$ equal 4 minutes.

The feedback time is calculated based on the maximum number of patients that can be served by the processing servers at each candidate node, $MaxP$ and the minimum shared link capacity between the candidate locations of the processing servers at the access layer to the LTE base station (i.e. uplink between ONU and OLT), $Cf_{min}$. The $MaxP$ is given as:

$$MaxP = N\ Pat \qquad (97)$$

where $N$ is the number of processing servers that can be connected at each candidate node due to the limited spaces at fog nodes and the complexity of the model to have more base stations. As the minimum link capacity will be shared by the maximum number of patients, the processing servers can serve at a node, the link capacity is divided by $MaxP$, to obtain the data rate for each patient to transmit the analysed data to the clinics ($\delta f$) as follows:

$$\delta f = Cf_{min}/MaxP \qquad (98)$$

The reason for limiting the feedback data rate by the data rate available for healthcare applications in GPON links is to increase the feedback data rate which, in turn, gives more time to transmit the raw ECG signal (252.8 kbits). This high available time to transmit the raw ECG signal uses a lower data rate. Therefore, fewer BSs will be activated. Note that, activating fewer BSs for a longer time is more efficient than activating a large number of BSs for a shorter time. This is because the idle power consumption of a BS is 63% of its total power. In this study, an LTE-M base station with the QPSK modulation scheme is considered which gives a minimum of 336 bps per physical resource block (PRB). Therefore, the number of PRBs for each patient to send the feedback data is given as:

$$Rf = \lfloor \delta f/336\ bits \rfloor \qquad (99)$$

where $Rf$ is the maximum number of PRBs allocated for each patient while ensuring that the given data rate does not exceed the maximum link capacities dedicated to the healthcare application. Hence, the data rate for feedback is calculated as below:

$$\delta b = Rf\ 336bps \qquad (100)$$

while the feedback time is:

$$\tau b = \alpha/\delta b \qquad (101)$$

Therefore, the remaining time to send the raw health data to the processing servers is given as:

$$\tau max = \tau t - \tau m - \tau b - \tau p \qquad (102)$$

hence a minimum data rate to transmit the raw health data to the processing server is:

$$\delta p = \Pi/\tau max \qquad (103)$$

As the data traversed the LTE base station, the number of PRBs that could be assigned to each patient to transmit his/her raw health data is calculated as below:

$$Rp = \lfloor \delta p/336\ bps \rfloor \qquad (104)$$

where $Rp$ was the maximum integer value is to ensure that the system could work within 4 minutes. Therefore, the data rate to send raw health data to the processing server is given as:

$$\delta a = Rp\ 336\ bps \qquad (105)$$

while the transmission time to send raw health data is calculated as below:

$$\tau a = \Pi/\delta a \qquad (106)$$

Meanwhile, the data rate to send the analysed health data at each processing server to the cloud storage for permanent storage was determined by dividing the minimum shared uplink or node capacity provisioned by a health M2M application from the processing server to the cloud storage, $Cs_{min}$ and the maximum number of patients that can be served at each candidate node, $MaxP$ as given below:

$$\delta c = Cs_{min}/MaxP \qquad (107)$$

while the time required to send the analysed health data to the cloud storage is calculated as below:

$$\tau c = \alpha/\delta c \qquad (108)$$

There are five approaches considered with 20%, 40%, 60%, 80% and 100% of the total number of patients in the 16 clinics, to investigate the impact of increasing the number of patients on the energy consumption of networking equipment and processing. This was done while considering the two protection scenarios (i.e. server protection and server and network protection scenarios) and the different number of allowed processing servers at each candidate node, $N$. Table 4 shows the data rate and transmission time to transmit the raw ECG data to the processing server, to transmit the analysed ECG data to the clinics and cloud for feedback and permanent storage, respectively. This is shown for the different number of processing servers per candidate node, $N$. Note that, the data rate and the time to transmit the raw ECG data to the processing servers for each number of processing servers per candidate node are the same for all approaches. This is because the data rate given to each patient is based on the number of allocated PRBs while ensuring the total data rate provided by the total number of PRBs per patient, is equal to or higher than the minimum data rate required so that the system can work within 4-minutes. Therefore, the same amount of PRBs are given to each patient under the different number of processing servers per candidate node, although their required minimum data rate is different.

Table 4: Data rate and related time for a different number of processing servers per candidate node $N$, for ECG monitoring applications

| Type of Data | 3 PSs | 4 PSs | 5 PSs | 6 PSs | 7 PSs | 8 PSs |
|---|---|---|---|---|---|---|
| Data rate to transmit ECG signal to processing server, $\delta a$ (kbps) | 1.344 | 1.344 | 1.344 | 1.344 | 1.344 | 1.344 |
| Transmission time to transmit ECG data to processing server, $ta$ (s) | 188.1 | 188.1 | 188.1 | 188.1 | 188.1 | 188.1 |
| Data rate to transmit analysed ECG data to clinics, $\delta b$ (kbps) | 1.008 | 0.672 | 0.672 | 0.336 | 0.336 | 0.336 |
| Transmission time to transmit analysed ECG data to clinics, $tb$ (s) | 0.254 | 0.381 | 0.381 | 0.762 | 0.762 | 0.762 |
| Data rate to transmit analysed ECG data to cloud storage, $\delta c$ (kbps) | 1.28 | 0.96 | 0.768 | 0.64 | 0.548 | 0.48 |
| Transmission time to transmit analysed ECG data to cloud storage, $tc$ (s) | 0.2 | 0.267 | 0.333 | 0.4 | 0.467 | 0.533 |

*5) Energy consumption model*

The maximum power consumption and the maximum capacity of the networking equipment and the processing server to calculate both idle power and load-dependent power are retrieved from datasheets and references. As for ONU, to obtain its energy proportional, the maximum capacity, $CONU$, was considered as the summation of maximum uplink capacity, 1.25 Gbps [42] and maximum downlink capacity, 2.5 Gbps [42]. The idle power of BS, PS, and content server are obtained from datasheets and references in [24], [43], and [44], respectively while the idle power for the other networking devices was considered to be 90% of the power consumption at maximum utilisation [24],[45], and [46]. As discussed before, the healthcare application is considered to contribute to 0.3% of the idle power of the shared devices. However, as the LTE-M BS shares capacity, antenna, radio, and hardware with the legacy LTE networks (20MHz) [47], and 6% allocation of healthcare application from the total M2M application [38] is supported by LTE-M network, therefore, the fraction of idle power contributed by the healthcare applications for BS is considered to be 0.42%. Note that, we also estimated that 6% allocations of the total PRBs are dedicated for healthcare application (i.e. 360 PRBs per second) as the LTE-M serves numerous types of M2M applications. Moreover, due to cooling and other overheads in the network devices, such as continuous power system at network sites; the power usage effectiveness (PUE) for IP over WDM, metro, and access networks of 1.5 is considered [48], [49]. The PUE for small distributed clouds applied in this work is 2.5 [22]. Also, a PUE of 2.5 was set for fog. Table 5 depicts the input parameters of the models for network architecture.

Table 5: Input parameters for networking and computing devices

| Parameter | Value |
|---|---|
| Maximum power consumption of core router (CRS-3), $PCR$ | 12300 W [45] |
| Core router capacity (CRS-3), $CCR$ | 4480 Gbps [45] |
| Maximum power consumption of cloud switch (Catalyst 6509), $PCLS$ | 2020 W [45] |
| Cloud switch capacity (Catalyst 6509), $CCLS$ | 320 Gbps [45] |
| Maximum power consumption of cloud router (7609), $PCLR$ | 4550 W [45] |
| Cloud router capacity (7609), $CCLR$ | 560 Gbps [45] |
| Maximum power consumption of content server, $PCS$ | 380.8 W [44] |
| Idle power consumption of content server, $ICS$ | 324.82 W [44] |
| Content server capacity, $CCS$ | 1.8 Gbps [44] |
| Maximum power consumption of cloud storage, $PCST$ | 4900 W [48] |
| Cloud storage capacity $CCST$ | 75.6 TB [48] |
| Maximum power consumption of aggregation router (7609), $PAR$ | 4550 W [24], [45] |
| Aggregation router capacity (7609), $CAR$ | 560 Gbps [24], [45] |
| Maximum power consumption of centre aggregation switch, (Catalyst 6509), $PCAS$ | 1766 W [45] |
| Centre aggregation switch capacity (Catalyst 6509), $CCAS$ | 256 Gbps [45] |
| Maximum power consumption of OLT, $POLT$ | 20 W [50] |
| OLT capacity, $COLT$ | 128 Gbps [50] |
| Maximum power consumption of ONU, $PONU$ | 8 W [42] |
| ONU capacity, $CONU$ | 3.75 Gbps |
| Maximum power consumption of LTE Base Station, $PBS$ | 528 W [51] |
| Idle power consumption of LTE Base Station, $IBS$ | 333 W [51] |
| Maximum power consumption of processing server, $PPS$ | 180 W [43] |
| Idle power consumption of processing server, $IPS$ | 78 W [43] |
| IP over WDM, access and metro network PUE, $\eta$ | 1.5 [48], [49] |
| Cloud and fog PUE, $c$ | 2.5 [22] |

## V. Performance Evaluation of the MILP model for ECG monitoring applications

This section presents the results and the analysis of the MILP model for the three scenarios: protection for the server without geographical constraints, protection for the server with geographical constraints and both protection for the server and network with geographical constraints and link and node disjoint, respectively. AMPL software with CPLEX 12.8 solver running on a high-perforamnce computing (HPC) cluster with a 12 core CPU and 64GB RAM was use as the platform to solve the MILP model. The performance of each scenario is investigated based on the percentage of patients considered in the network and the number of processing servers that can be served at each candidate node. Also, the energy penalty of networking equipment with an increasing level of resilience is evaluated.

### 1) Protection for servers without geographical constraints

In this section, the performance of the non-resilient scenario is used as a benchmark to evaluate the resilient scenario without geographical constraints in terms of the energy consumption of networking equipment and processing for ECG monitoring applications.

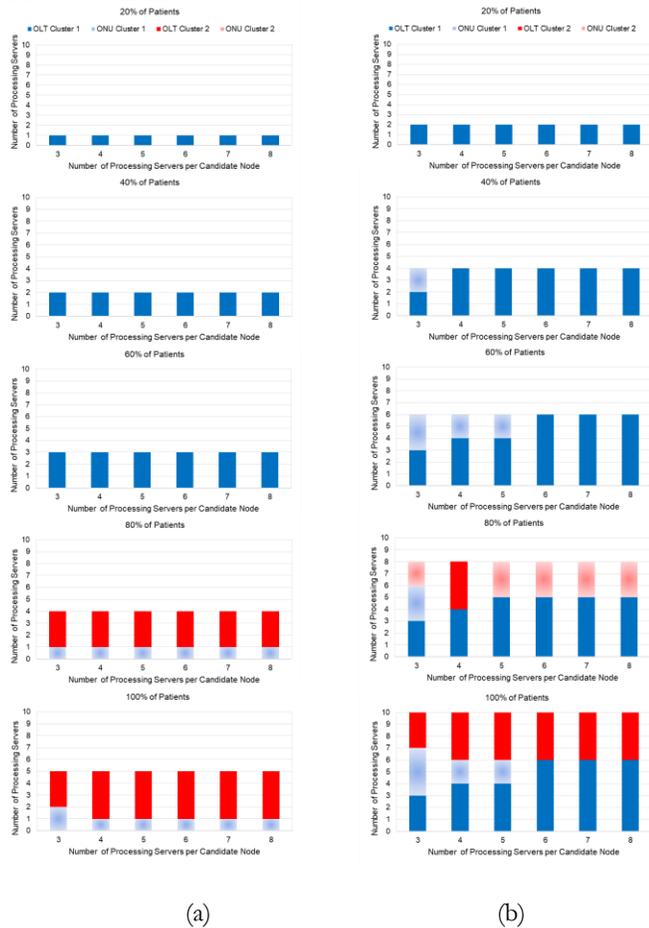

(a)          (b)

Figure 2: Optimal location of processing servers for (a) non-resilient scenario and (b) resilient scenario without geographical constraints

The results in Figure 2, show that the number of processing servers for the resilient scenario is double that of the non-resilient scenario. This is because, the non-resilient scenario only has primary processing servers while the resilient scenario consists of a secondary processing server for each primary processing server, for server protection purposes. The results show that increasing the percentage of patients, has resulted in increasing the number of processing servers. For the non-resilient scenario, at the demand level of 20%, 40%, 60%, 80% and 100%, the number of processing servers required to serve all patients are one, two, three, four and five, respectively. Meanwhile, for the resilient scenario, the number of processing servers for each demand level is double that of the number of processing server of the non-resilient scenario. Figure 2 also shows that the OLT is always chosen to place the processing servers as it is the nearest shared point to the patients (i.e. the OLT is connected to all base stations of the same cluster) which reduces the number of required processing servers and the number of hops to transmits the ECG signal to the processing servers.

The processing servers are placed at only one cluster when the percentage of patients considered in the network is equal to or less than 60% as shown in Figure 2-(a) and Figure 2-(b). This is because all patients can be served by the base stations in one cluster only. Therefore, for the resilient scenario without geographical constraints, to reduce the number of utilised networking equipment in the network, the ONU is selected to place the remaining processing servers, which cannot be allocated at the OLT at the same cluster, while for the non-resilient scenario the processing servers are only placed at the OLT.

However, increasing the percentage of patients to 80% and 100% has resulted in utilising the BSs, ONUs and OLTs in both clusters. For the non-resilient scenario, the primary processing servers are placed at the OLT and ONU of different cluster when the demand increases to 80% and 100%. This is because the OLT does not have enough capacity to support all of the traffic. The OLT of cluster 2 is occupied first, and the remaining demands are sent to the ONU of the cluster 1, to reduce the total amount of data traversing the network as ONUs are directly connected to the patients. For the resilient scenario without geographical constraints, at a demand level of 80%, and three PSs available at each candidate node, the OLT and ONUs of cluster 1 are occupied first, and the remaining demands are sent to the ONU of cluster 2. This is due to the same reason, as explained for the non-resilient scenario. However, when the demand level increases to 100%, the OLTs of both clusters and only the ONU of one cluster are used. The model did not use multiple ONUs to place the processing server to reduce the number of utilised Ethernet switches. Also, when five processing servers are allowed at each candidate node, the processing servers are placing at both OLTs and one ONU mainly to reduce the number of utilised base stations, as the base station consumed more energy than the Ethernet switch.

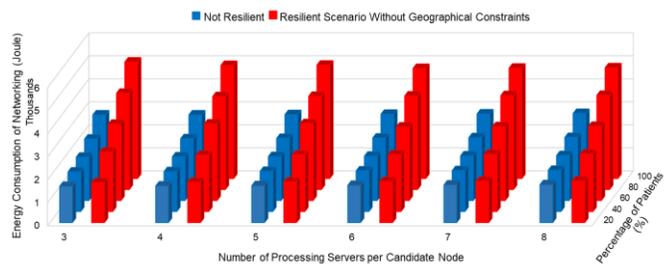

Figure 3: Energy consumption of networking equipment for non-resilient scenario and resilient scenario, without geographical constraints

The results in Figure 3, show that the increasing rate of energy consumption of networking equipment due to the increasing demand for the resilient scenario is higher than the non-resilient scenario. The results also show that the energy consumption of networking equipment of the resilient scenario without geographical constraints is always higher than the non-resilient scenario for all levels of demand and number of processing servers per node. This is because, the total traffic traversing the networking equipment for the resilient scenario is double compared to the non-resilient scenario, hence increasing the total number of utilised networking equipment. This increase in energy consumption is one of the key penalties for having resilience.

Figure 3 also shows that at a demand level equal to or more than 40%, the energy consumption of networking equipment of the resilient scenario reduced significantly when the number of processing servers increased from three to eight. This is because increasing the number of processing servers per candidate node has resulted in placing the processing servers at their optimal locations besides reducing the number of utilised nodes.

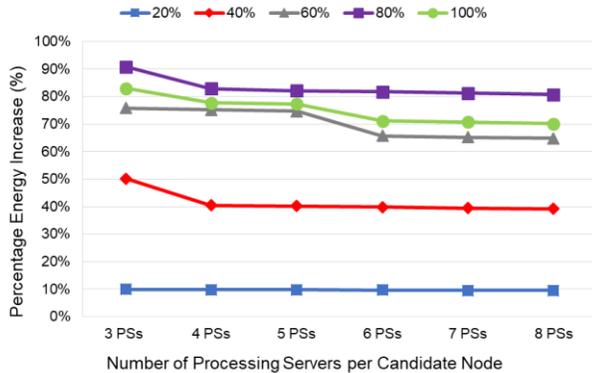

Figure 4: Percentage of energy penalty of networking equipment for resilient scenario, without geographical constraints compared to non-resilient scenario

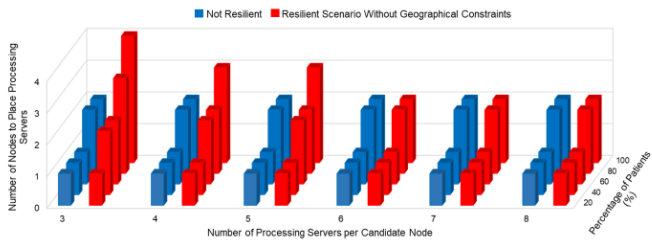

Figure 5: Number of nodes used to place processing servers for non-resilient scenario and the resilient scenario, without geographical constraints

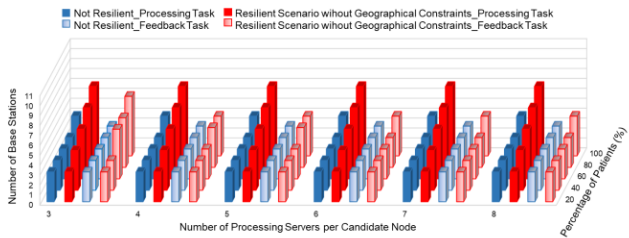

Figure 6: Number of base stations used to send the raw ECG signal for processing and the analysed ECG signal for feedback, for non-resilient scenario and resilient scenario without geographical constraints under different percentages of patients and number of processing servers per candidate node

The results in Figure 4, show that the energy penalty (defined as the difference in energy consumption between the resilient and the non-resilient cases) increases when the level of demand increases from 20% to 80%. This is because, at demand levels of 20% to 80%, the number of utilised base stations to serve all patients to send their ECG signal to the processing servers for the non-resilient scenario are the same while for the resilient scenario, the number of base stations increases with the increasing demand, as shown in Figure 6. The increasing number of base stations under the resilient scenario is because, each patient will send two ECG signals to both primary and secondary processing servers, hence requiring a high number of base stations to serve all patients and this number increases as the demand increases.

For the non-resilient scenario, each patient only sends one ECG signal to the primary processing servers, and the same number of base stations are used, as they can accommodate the increasing demand by up to 80%. However, at a demand level of 100%, the energy penalty is lower than 80%. This is because, at a demand level of 100%, the number of base stations used for the non-resilient scenario increases, hence increasing the energy consumption of networking equipment of the non-resilient scenario. Figure 4, also shows that increasing the number of processing servers at each candidate node can significantly reduce the energy penalty when the demand is equal to or is higher than 40%. This is due to the reduction in the number of nodes used to place the processing servers for the resilient scenario as shown in Figure 5, where more processing servers can be placed at the same node when the number of processing servers allowed at each candidate node increases.

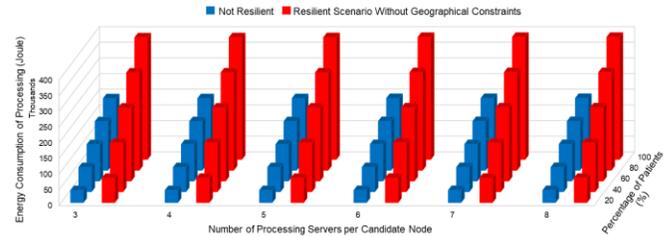

Figure 7: Energy consumption of processing for non-resilient scenario and resilient scenario, without geographical constraints

The results in Figure 7, show that the energy consumption of processing for the resilient scenario is higher than the non-resilient scenario. This is because the number of processing servers for the resilient scenario is double that of the non-resilient scenario. The results also show that the energy consumption of processing increases, as the demands increase for both scenarios. This is because increasing the number of patients increases the number of processing servers proportionally. However, the same total number of servers is used in both scenarios under constraints on the number of processing servers per candidate node, as the patients were optimally consolidated in the servers. Also, for both scenarios, there is a slight increase in energy consumption, when more processing servers are allowed per candidate node. The increase is due to the increasing utilisation time of the processing servers to send the feedback and storage traffic, with the increasing number of processing servers per candidate node, as shown in Table 4.

### 2) Protection for servers with geographical constraints

In this section, the performance of the resilient scenario without geographical constraints is used as a benchmark to evaluate the increasing level of resilience gained by considering the geographical constraints, in terms of the energy consumption of networking equipment and processing.

The results in Figure 8, show that the OLT is always used to place the processing servers as in the previous scenarios. The results also show that the processing servers are placed at only one cluster, when the percentage of patients is equal to or less than 60%. This is to reduce the utilisation of the networking equipment. However, due to the geographical constraints, at least two locations are required to place the primary and secondary processing servers. Therefore, both OLT and ONU of the same clusters are selected to place the processing servers, separately.

Figure 8 also shows that at a high level of demand (i.e. 80% and 100%), the BSs, ONUs and OLTs from both clusters are utilised. The results show that at a demand level of 80% for all processing servers per candidate node, the OLT and ONUs of cluster 1 are occupied first, and due to the limited number of resources of the base stations in cluster 1 to serve the patients, the remaining demand is sent to the ONU of cluster 2. This is to reduce the total amount of data traversing the network as ONUs are directly connected to the patients. The results also show that, at the demand level of 80%, when the number of processing servers allowed at each node increases to four, the number of utilised nodes to place the processing servers are reduced as more processing servers are placed at the OLT.

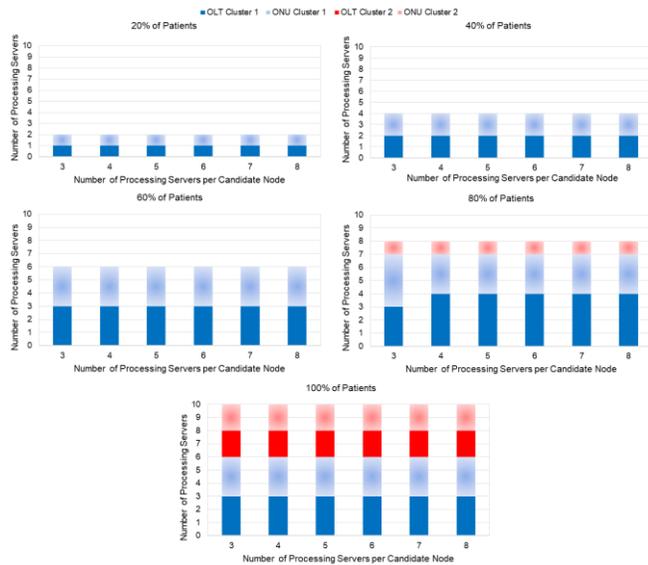

Figure 8: Optimal location of processing servers for resilient scenario, considering geographical constraints

However, when the demand level increases to 100%, the OLT and the ONU of both clusters are used to accommodate the increasing number of processing servers in the network for all processing servers per candidate node. The results show that increasing the number of processing servers per candidate node does not affect the location of placing the processing servers, as optimal locations are selected.

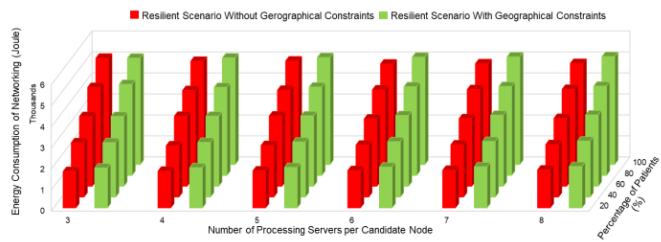

Figure 9: Energy consumption of networking equipment for resilient scenario, without geographical constraints and resilient scenario considering geographical constraints

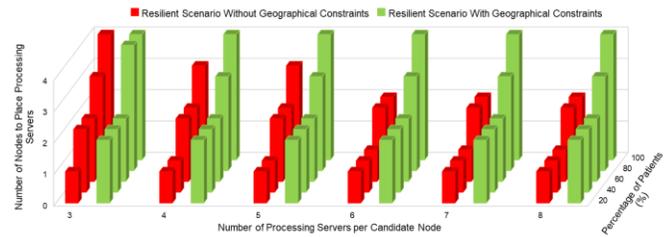

Figure 10: Number of nodes used to place processing servers for resilient scenario, without geographical constraints and resilient scenario considering geographical constraints

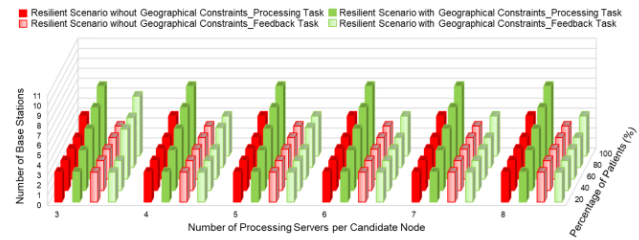

Figure 11: Number of base stations used to send the raw ECG signal for processing and the analysed ECG signal for feedback, for resilient scenario without geographical constraints and resilient scenario, with geographical constraints under different percentages of patients and number of processing servers per candidate node

The results in Figure 9 show that at demand levels of 40%, 60% and 100% and when the number of available processing servers is three, the energy consumption of networking equipment for both scenarios is the same. This is due to the same number of utilised networking equipment, where the same number and location of nodes are used to place the processing servers, and the same number of base stations are used to serve the patients to send their ECG signal to the processing servers, for both scenarios, as shown in Figure 10 and Figure 11, respectively.

However, at a demand level of 60% and when four and five processing servers are allowed at each candidate node, the energy consumption of networking equipment with the more resilient scenario is slightly higher than the resilient scenario without geographical constraints, although the same number of base stations and nodes are used to place the servers for both scenarios. This is due to the different locations of the processing servers in the network for both scenarios, where for the more resilient scenario, the location of processing servers has led to more data traversing the networking equipment than the resilient scenario, without geographical constraints.

Meanwhile, for the other levels of demand and number of processing servers per candidate node, the energy consumption of the more resilient scenario is higher than the resilient scenario without geographical constraints, as shown in Figure 9. This is because, considering the geographical constraint increases the total number of utilised nodes to place the processing servers, as shown in Figure 10. Hence, the number of utilised networking equipment in the more resilient scenario increases. This increase in energy consumption is the penalty for having a higher level of resilience.

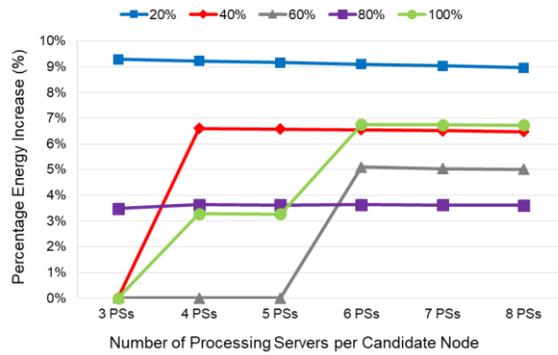

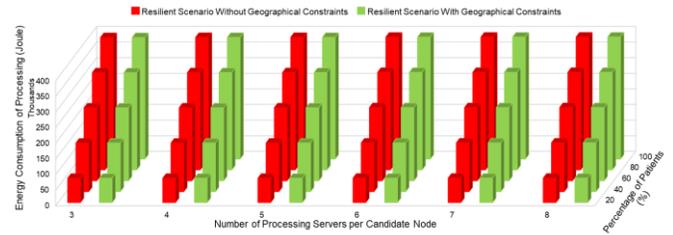

Figure 12: Percentage of energy penalty of networking equipment for the resilient scenario, considering geographical constraints, compared to the resilient scenario without geographical constraints

The results in Figure 12, show that increasing the level of resilience to consider geographical constraints, does not incur any energy penalty at demand levels of 40%, 60% and 100%; when three processing servers are available at each node. This is due to the same number of utilised networking equipment in both scenarios (i.e. nodes to place the processing servers and base stations to send the processing traffic). However, at demand levels of 20% and 80%, increasing the level of resilience to consider geographical constraints has resulted in an energy penalty. This is because, at these specific demands, a higher number of nodes are used to place the processing servers for the more resilient scenario, compared to the resilient scenario without geographical constraints, as shown in Figure 10.

Figure 12 also shows that increasing the number of allowed processing servers at each candidate node can increase the energy penalty when the demand is more than 20%. The increase in energy penalty is due to the decreasing number of nodes available to place the processing servers with a resilient scenario without geographical constraints, as shown in Figure 10. However, at demand levels of 20% and 80%, increasing the number of processing servers per candidate node does not result in a significant impact on the energy penalty. This is because, at this specific demand, the same number of nodes are used to place the processing servers and the same number of base stations are used to send the ECG signal to the processing servers in both scenarios, as shown in Figure 10 and Figure 11, respectively.

Figure 12 also shows that, when the number of processing servers allowed at each node is equal to or higher than six, the energy penalty decreases as the demand increases from 20% to 80%. This is because the same number of base stations are used in both scenarios to send the ECG signal to the processing servers, as shown in Figure 11. However, the energy penalty at a demand level of 100% is higher than 40%, as the number of nodes used to place the processing servers for the more resilient scenario is doubled, in comparison to the resilient scenario without geographical constraints, as shown in Figure 10.

Figure 13: Energy consumption of processing for resilient scenario without geographical constraints and resilient scenario considering geographical constraints

The results in Figure 13 show that, for both scenarios, the energy consumption of processing increases as the level of demand increases for all number of processing servers per candidate node. This is because increasing the demand increases the number of processing servers proportionally. For all processing servers allowed at each candidate node, the energy consumed is equal for both resilience levels. This is because the same number of servers will be utilised regardless of their location, as the patients are optimally consolidated in the servers. Also, there is a slight increase in the energy consumption of processing, when the number of processing servers per candidate node increases. This is due to increasing the utilisation time of the processing servers to send the feedback and storage traffic under the increasing number of processing servers per candidate node, as explained previously.

### 3) Protection for servers with geographical constraints and network with link and node disjoints

In this section, the performance of the resilient scenario with geographical constraints is used as a benchmark to evaluate the increased level of resilience (in disjoint link and node resilience) in terms of the energy consumption of networking equipment and processing for ECG monitoring applications.

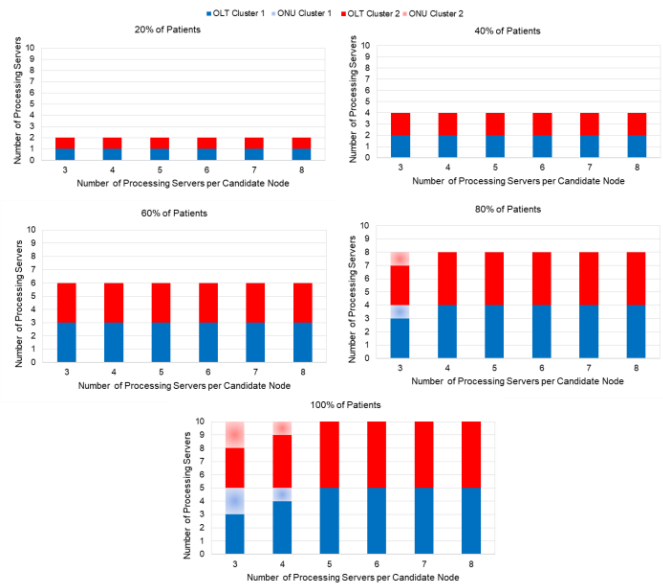

Figure 14: Optimal location of processing servers for the resilient scenario considering the geographical constraint for server protection and link and node disjoint resilience for network protection

The results in Figure 14 show that the processing servers are only placed at the (optical line terminals) OLTs in both clusters when the number of processing servers allowed at each candidate node is equal to or greater than the total number of primary or secondary processing servers required in the network. This is for two reasons. The first is to reduce the number of nodes (i.e. Ethernet switches) used to place the processing servers, as the OLTs are the nearest shared point to the patients. The second is because each cluster is used to place the same set of processing servers. For instance, cluster 1 is used to place only primary processing servers, while cluster 2 is used to place only secondary processing servers. Therefore, when the number of processing servers allowed at each candidate node is less than the number of primary and secondary processing servers required, the (optical network units) ONUs in both clusters are utilised to place the remaining processing servers under increasing demands.

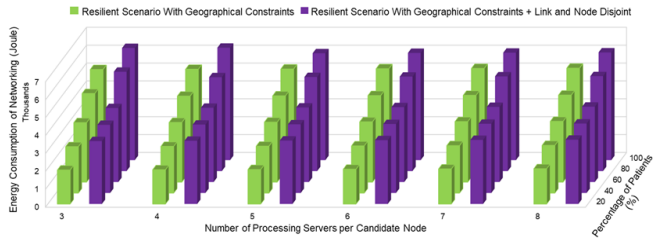

Figure 15: Energy consumption of networking equipment for the resilient scenario considering the geographical constraints and the resilient scenario with geographical constraints and link and node disjoint resilience

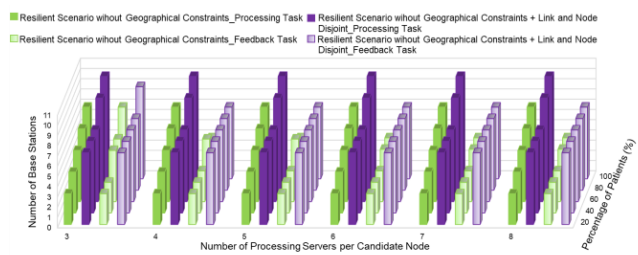

Figure 16: Number of base stations used to send the raw ECG signal for processing and the analysed ECG signal for feedback, for the resilient scenario considering the geographical constraints and the resilient scenario considering geographical constraints and link and node disjoint under different percentages of patients and number of processing servers per candidate node

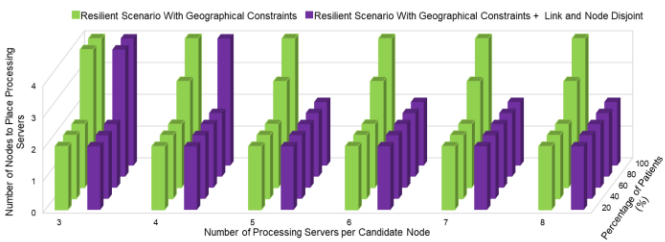

Figure 17: Number of nodes used to place the processing servers for the resilient scenario considering the geographical constraints and the resilient scenario with geographical constraints and link and node disjoint resilience

The results in Figure 15 show that the energy consumption of networking equipment for both scenarios increases as the demand increases for all the different numbers of processing servers per candidate node considered. This is due to the increasing amount of traffic in the network, hence increasing the total number of networking equipment utilised in the network. The results also show that, for all levels of demands and number of processing servers per candidate node, the energy consumption of networking equipment for the more resilient scenario is always higher than the resilient scenario that only considers geographical constraints. This is due to the high number of base stations utilised in the more resilient scenario, as shown in Figure 16. Note that, each base station is connected to only one OLT in the network. Therefore, considering disjoint links and nodes for network protection has increased the number of base stations without maximising the utilisation of their resources to send the processing traffic to both primary and secondary processing servers.

It is worth noting that the number of nodes used to place the processing servers at demand levels of 80% and 100% in the more resilient scenario is lower than the resilient scenario with geographical constraints when the number of processing servers per candidate node is equal to or more than four and five, respectively, as shown in Figure 17. However, as the energy consumed by a single base station is approximately 1.5x higher than the energy consumed by a single node (i.e. Ethernet switch) to place the processing servers, therefore there is an energy penalty with the link and node disjoint resilience scenario.

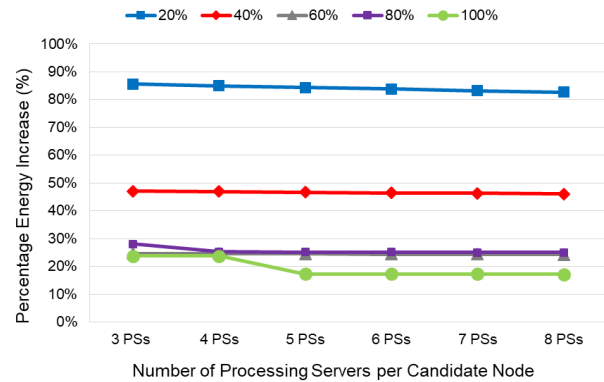

Figure 18: Percentage energy penalty of networking equipment for the resilient scenario considering the geographical constraints and link and node disjoint resilience compared to the resilient scenario considering the geographical constraints

The results in Figure 18 show that the energy penalty with the link and node disjoint resilience scenario decreases as the demand level increases from 20% to 60% and 80% to 100%. This is because the total number of base stations utilised in the resilient scenario, that only consider the geographical constraint, increases with the increases in demand in the network, as shown in Figure 16. This increases the energy consumption of networking equipment for the resilient scenario that only considers the geographical constraint as the demand levels increase. However, at a demand level of 60%, the energy penalty is lower than at a demand level of 80%. This is because, at a demand level of 80%, the number of base stations used for the more resilient scenario starts to increase, hence increasing the energy consumption of networking equipment of the more resilient scenario.

Figure 18 also shows that, at demand levels of 80% and 100%, increasing the number of processing servers per candidate node to 4 and 5, respectively, decreases the energy penalty. This is because the number of nodes (i.e. Ethernet switches) used to place the processing servers for the more resilient scenario reduces while the same number of nodes are used for the resilient scenario that only considers geographical constraints as shown in Figure 17.

The results in Figure 19 show that the energy consumption of processing for both scenarios increases with the increasing level of demand. This is due to the increased number of utilised processing servers in the network, as explained previously. Figure 19 also shows that increasing the level of resilience does not increase the energy consumption of processing. Also, the energy consumption of processing has slightly increased with the increasing number of processing servers per candidate node. The same energy for processing in both scenarios and the increase in energy for processing in both scenarios with the increase in the number of processing servers per candidate node is for the same reason as explained previously.

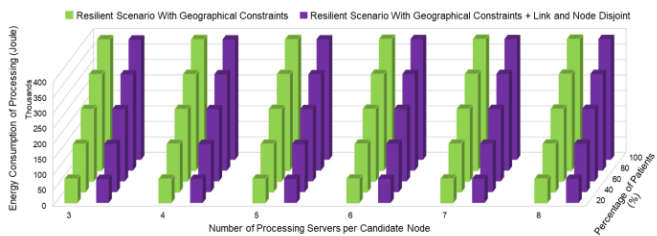

Figure 19: Energy consumption of processing for the resilient scenario considering the geographical constraints; and the energy consumption of the resilient scenario with geographical constraints and link and node disjoint resilience

## VI. Heuristics Models

*1) Energy optimised resilient infrastructure fog computing without geographical constraints (EORIWG) heuristic*

The EORIWG heuristic determines the BSs to be used to serve patients to send raw health data and receive feedback data and the nodes to place primary and secondary processing servers at the access network so that the energy consumption of both networking and processing are minimised. Figure 20 shows the flow chart of the EORIGW heuristic.

The heuristic begins by grouping the clinics based on the number of BSs in cluster 1 it can connect to and sorts the groups in ascending order. For each group, the clinics are sorted based on the total number of BSs in both clusters the clinic can connect to in ascending order. The heuristic assigns first the clinic with the smallest number of connections to the BSs in cluster 1 and the smallest number of connections to all BSs in both clusters, to the BSs to help in reducing the utilisation of OLTs. Also, it ensures that all clinics are assigned to BSs.

The assignment of clinic patients to a BS is as follows: The heuristic sorts the BSs that have a connection to the clinic under consideration starting with BSs previously used by the healthcare application that has available resources. These BSs are sorted in ascending order based on the total number of clinics the BS can serve followed by the unused BSs in cluster 1 in descending order and followed by the unused BSs in cluster 2 also in descending order. Sorting the activated BSs in ascending order is used to reduce the number of utilised BSs while the descending order of unused BSs in cluster 1 followed by the unused BSs in cluster 2 is used to ensure that options are left open until late in the allocation process while minimising the utilisation of the OLTs. Then, the patients of the clinic under consideration are consolidated to the minimum number of BSs to reduce the number of BSs used by the healthcare application. Note that, for each patient, the heuristic assigned double resources to clinics so that they send their health data to both primary and secondary processing servers.

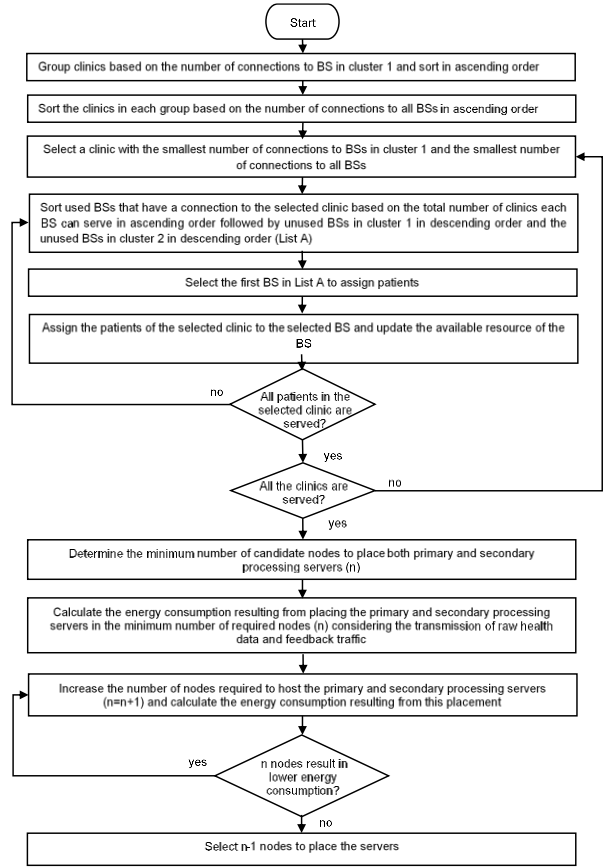

Figure 20: Flow chart for EORIGW heuristic

The heuristic then determines the number of primary and secondary processing servers required to serve the patients and the nodes hosting them. The candidate nodes to host the servers are the ONUs connected to the BSs selected to serve the patients and the OLTs. Considering the minimum number of nodes required to host both primary and secondary servers to serve all the patients (which is based on the maximum number of servers a node can host), the heuristic finds the combination of candidate nodes to host the primary and secondary processing servers that result in minimum energy consumption. Limiting the number of nodes to place the primary and secondary processing servers reduces the utilisation of the Ethernet switches to serve the processing servers.

The energy consumption that results from hosting both primary and secondary processing servers at a combination of candidate nodes is calculated by routing the traffic (raw health data traffic) to the nearest node with available processing capacity of the combination of candidate nodes under consideration based on minimum hop routing.

Also, BSs to send feedback traffic from combination of candidate nodes to clinics are selected using the same approach used to select BSs to send raw health data. Note that BSs different from those used to send raw health data are used to send feedback traffic. The difference is because the size of the analysed data for feedback is smaller than the raw health data.

The combination of nodes hosting servers considering the minimum number of nodes required to host primary and secondary servers to serve all the patients that result in minimum energy consumption is selected. The heuristic increases the number of candidate nodes to host servers and repeats the above process. If the energy consumption resulting from using this combination of nodes is lower than the energy consumed with combination of nodes hosting servers considering the minimum number of nodes required to host servers, the heuristic examines placing servers in more candidate nodes. Else, the minimum number of nodes required to host servers is selected to place servers.

### 2) Energy optimised resilient infrastructure fog computing with geographical constraints (EORIG) heuristic

The EORIG heuristic determines the BSs to serve patients so as to send raw health data and receive feedback data and the nodes to place primary and secondary processing servers at the access network so that the energy consumption of both networking and processing is minimised and the primary and secondary servers are node disjoint (geographical constraints). Below is the list of the changes made for the EORIG heuristic compared to EORIGW heuristic:

1. The number of nodes to place processing servers is based on the total number of nodes to place primary and secondary processing servers in disjoint nodes.
2. Assigning patients from BSs to the primary processing servers is done first and the nodes used to place the primary processing servers are removed from the combination of nodes before assigning the same patients from the BSs to the secondary processing servers.

### 3) Energy optimised resilient infrastructure fog computing with geographical constraints and link and node disjoints (EORIGN) heuristic

As in previous heuristics, the EORIGN heuristic determines the BSs to be used to serve the patients so as to send the raw health data and receive feedback data. It also determines the nodes to be used to place the primary and the secondary processing servers at the access network so that the energy consumption of both networking and processing are minimised. Figure 21 shows the flow chart of the EORIGN heuristic.

In the EORIGN heuristic, the selection of the locations to host the primary servers and secondary servers are done separately to ensure that the traffic to the primary server and the traffic to the secondary servers are routed separately (link and node disjoint). In this process, the heuristic begins by selecting a cluster to assign the patients in the clinics to the primary processing server. Then the heuristic groups the clinics based on the number of BSs in the selected cluster it can connect to and sorts the groups in ascending order. For each group, the clinics are sorted based on the total number of patients it serves in ascending order. The heuristic assigns first the clinic with the smallest number of connections to the BSs in the selected cluster and the smallest number of patients it serves to the BSs to ensure each clinic can be served by at least one BS and to help in packing the BSs (packing is optimum when equipment have high idle power consumption).

The assignment of clinic patients to a BS is as follows: The heuristic sorts the BSs in the selected cluster that has a connection to the clinic under consideration starting with BSs previously used by the healthcare application that has available resources. These BSs are sorted in ascending order based on the total number of clinics the BS can serve followed by the unused BSs in the selected cluster in descending order. The ascending order of activated BSs is used to reduce the number of utilised BS while the descending order of unused BS in the selected cluster is used to ensure that options are left open until late in the allocation process. Then, the patients of the clinic under consideration are consolidated in the minimum number of BSs to reduce the number of BSs used by the healthcare application.

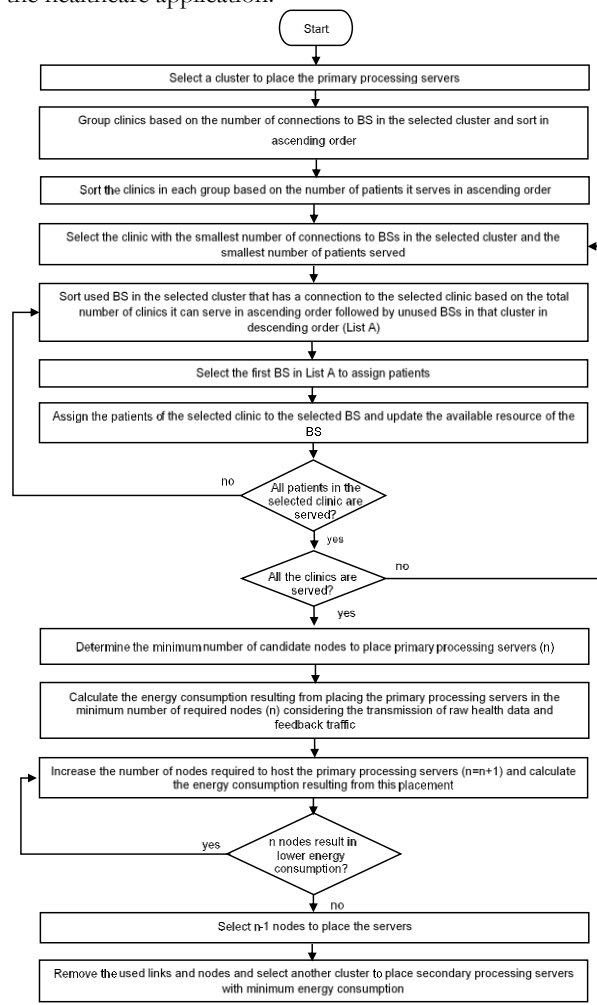

Figure 21: Flow chart of EORIGN heuristic

The heuristic then determines the number of primary processing servers required to serve the patients and the nodes hosting them. The candidate nodes to be used to host the servers are the ONUs connected to the BSs selected to serve the patients and the OLT of the selected cluster. Considering the minimum number of nodes required to host servers to serve all the patients (which is based on the maximum number of servers a node can host), the heuristic finds the combination of candidate nodes to host the primary processing servers that result in minimum energy consumption. Limiting the number of nodes used to place the primary processing servers is for the same reason as explained for EORIWG which is to reduce the number of Ethernet switches used to serve the processing servers.

The energy consumption that results from hosting servers at a combination of candidate nodes in the selected cluster is calculated as explained for EORIWG heuristic. The BSs to be used to send feedback traffic from combination of candidate nodes to clinics are selected using the same approach used to select BSs to send raw health data. Note that different BSs are used to send raw health data and to send feedback traffic for the same reason as explained in EORIWG heuristic. The combination of nodes hosting servers considering the minimum number of nodes required to host primary processing servers to serve all the patients that result in minimum energy consumption is selected. As in EORIWG heuristic, the heuristic increases the number of candidate nodes used to host servers. The energy consumption resulting from using this combination of nodes is calculated and compared to the energy consumption resulting from the combination of nodes hosting servers considering the minimum number of nodes required to host servers. If the latter is lower, the heuristic examines placing servers in more candidate nodes. If the former is lower, the minimum number of nodes required to host servers is selected to place servers.

Next, the heuristic removes the links and nodes used to send the traffic to or from primary processing servers and selects another cluster to assign patients in the clinics to the secondary processing servers. Different clusters are used to host the primary and secondary processing servers, which is due to the link and node disjoint resilience mandated for network protection. The same process is used to allocate patients to the BSs to send raw health data and to receive analysed health data feedback. It is also used for the selection of locations to host the secondary processing servers and to determine the optimal location to host the secondary processing server.

## VII. RESULTS AND ANALYSIS OF THE HEURISTIC MODELS

In this section, we evaluate the performance of the developed heuristics for server protection, the EORIWG heuristic and EORIG heuristics, and heuristic for server and network protection, EORIGN heuristic, compared to the MILP results in term of the energy consumption of networking equipment and processing. The evaluations are performed for both ECG monitoring applications and fall monitoring applications considering all levels of demand and 100% of demand level, respectively. As in the previous chapters, the heuristics are running on a normal PC with 3.2 GHz CPU and 16 GB RAM.

### 1) Energy optimised resilient infrastructure fog computing without geographical constraints (EORIWG) heuristic

Figure 22 shows that the total energy consumption of EORIWG heuristic is equal to that of the MILP model when the demand levels are 20% and 40% for all number of processing server per candidate node. This is due to the ability to use the minimum number of primary and secondary processing servers and the number of nodes to place the processing servers that are built into the EORIWG heuristic while assigning the patients from clinics to the processing servers.

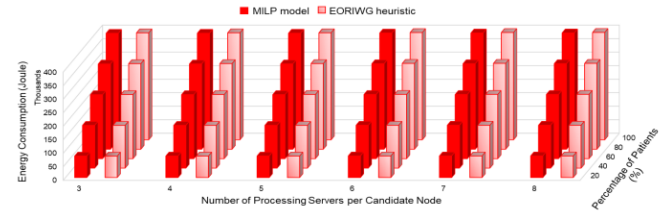

Figure 22: Total energy consumption of both networking equipment and processing for the MILP model and the EORIWG heuristic with different percentages of the total number of patients for different number of processing servers per candidate node

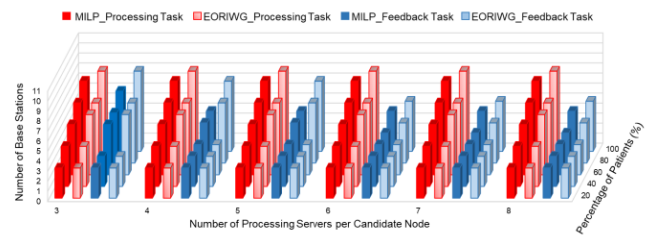

Figure 23: Number of base stations used to serve the processing and feedback tasks for the MILP model and the EORIWG heuristic with different percentages of the total number of patients for different number of processing servers per candidate node

Figure 22 also shows that the total energy consumption of the EORIWG heuristic is higher than the MILP model with an average of 0.17%, 0.42% and 0.44%, at demand levels of 60%, 80% and 100%, respectively. The higher energy consumed in the EORIWG heuristic is because at demand levels of 60% and 100%, increasing the patients has resulted in utilising more base stations to send the raw ECG data to the processing servers as shown in Figure 23. In the EORIGW heuristic, all base stations in cluster 1 are utilised, and due to the different connections of each clinic to the base stations, the utilisation of the resources in the selected base stations are not maximised. Therefore, the base stations in cluster 2 are also used to serve the patients from the remaining clinics.

Also, at demand levels of 80% and 100%, the number of base stations utilised in the EORIGW heuristic to send the feedback traffic is higher than in the MILP model, as shown in Figure 23, hence more networking equipment energy is consumed in the EORIWG heuristic compared to the MILP model. Note that, increasing the number of base stations to send the processing traffic results in more impact on the energy of networking equipment compared to the growing number of base stations used to send the feedback traffic. Also note that, in EORIWG heuristic, the number of nodes used to place the processing servers is equal to the minimum required nodes to place both primary and secondary processing servers. Therefore, due to the restricted number of nodes to place the processing servers, the centre aggregation switch (CAS) is activated in the EORIWG heuristic to send the ECG signal to the processing servers located at different clusters when the demand levels increase to or more than 60%. The utilisation of the CAS has increased the energy consumption of networking equipment in the EORIWG heuristic.

*2) Energy optimised resilient infrastructure fog computing with geographical constraints (EORIG) heuristic*

The results in Figure 24 show that the total energy consumption of EORIG heuristic is equal to that produced by the MILP model at demand levels of 20% and 40% for all number of processing servers per candidate node. This is mainly due to the ability to utilise the minimum number of primary and secondary processing servers and number of nodes to place the processing servers that are built in the EORIG heuristic while assigning the patients to the processing servers. Also, as the size of demand is small, the same number of networking equipment is utilised in both EORIG heuristic and MILP model.

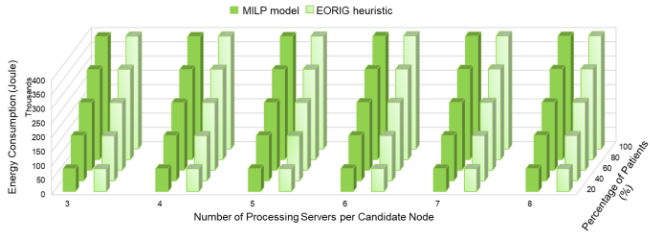

Figure 24: Total energy consumption of both networking equipment and processing for the MILP model and the EORIG heuristic with different percentage of patients for different number of processing servers per candidate node

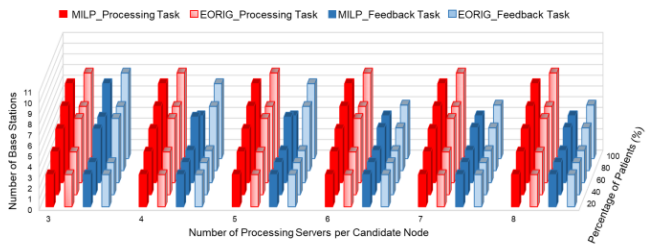

Figure 25: Number of base stations used to serve the processing and feedback tasks for the MILP model and the EORIG heuristic with different percentage of patients, for different number of processing servers per candidate node

Figure 24 also shows that the total energy consumption of the EORIG heuristic is higher than that produced by the MILP optimisation model with an average increase of 0.17%, 0.39% and 0.39% when the demand levels are 60%, 80% and 100%, respectively. The high energy consumed in EORIG heuristic at demand levels of 60% and 100% is due to the high number of utilised base stations to send the processing and feedback traffic as shown in Figure 25. Note that, the base stations in cluster 1 and cluster 2 are used to serve the processing traffic due to the limitation of the connection between the clinics and the base stations. Also, at 80% and 100% of the maximum demand level, the higher energy consumed in the EORIG heuristic is due to the utilisation of the centre aggregation switch (CAS) to relay the processing traffic between the clusters to the processing servers.

*3) Energy optimised resilient infrastructure fog computing with geographical constraints and link and node disjoints (EORIGN) heuristic*

The results in Figure 26 show that the total energy consumption of EORGN heuristic is equal to the energy consumption reported by the MILP optimisation model at demand levels of 20%, 40%, 60%, and 80% for all number of processing servers per candidate node. This is due to the same amount of utilised networking equipment and processing servers in both models. Figure 26 also shows that, at a demand level of 100%, the total energy consumption of the EORIGN heuristic is slightly higher than the MILP model with an average difference of about 0.1%. This is due to the limited number of connections between the base stations and the clinics in each cluster. Hence resulting in the utilisation of a higher number of base stations in the EORIGN heuristic, as shown in Figure 27 to serve the processing traffic without maximising the utilisation of its resources.

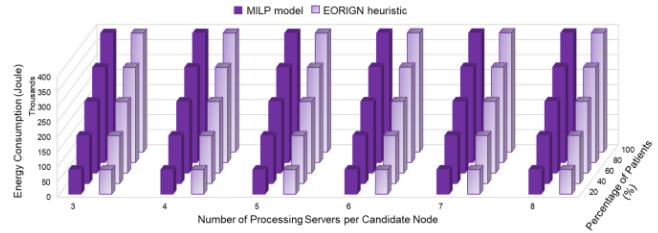

Figure 26: Total energy consumption of networking equipment and processing for the MILP model and the EORIGN heuristic with different percentages of the total number of patients for different number of processing servers per candidate node

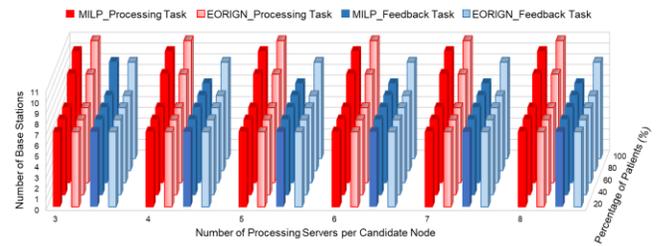

Figure 27: Number of base stations used to serve the processing and feedback tasks for the MILP model and the EORIGN heuristic with different percentages of the total number of patients for different number of processing servers per candidate node

## VIII. CONCLUSIONS

This work has investigated the impact of increasing the level of resilience on the energy consumption of networking equipment and processing to serve fog-based health monitoring applications. This is accomplished by considering three resilience scenarios. The first two scenarios are for server protection related to geographic location while the third scenario considers server and network protection with geographical constraint and link and node disjoint, respectively. The results reveal that increasing the level of resilience to consider resilience scenario without geographical constraints has increased the energy consumption of both networking equipment and processing compared to non-resilience scenario. This is mainly due to the high number of utilised networking equipment and processing servers as considering resilient resulted in doubling the amount of traffic to send the data to both primary and secondary processing servers. Meanwhile increasing the level of resilience to consider geographical constraint has resulted in the high energy penalty at low demand. This is because more nodes are utilised, to place the processing servers under the geographic constraints. However, when the demand level increases from 40% to 100%, increasing the level of resilience does not incur an energy penalty, and this depends on the number of processing servers allowed at each candidate node. Also, increasing the number of processing servers per candidate node at a demand level of more than 20%, can either decrease or increase the energy penalty. The increase in the energy penalty is because of the reduction in the number of nodes needed to place the processing servers in the resilient scenario, without geographical constraints. On the other hand, the decrease in the energy penalty is because of the reduced number of nodes needed to place the processing servers in the resilient scenario, with geographical constraints. However, the energy penalty due to considering geographical constraints at a demand level of more than 20%, is less than 7%. The results also show that the same energy of processing is consumed in both resilient scenarios, for all processing servers per candidate node. This is because the same number of servers are used in both scenarios, as the patients were optimally consolidated in the processing servers. Increasing the level of resilience to consider geographical constraint for server protection and link and node disjoint resilience for network protection compared to only geographic constraints gives the same energy consumption of processing, while increasing the energy consumption of networking equipment. The results indicate that considering additional disjoint link and node resilience has resulted in a low network energy penalty at high demands due to the activation of a large part of the network in any case due to the demands. Also, increasing the number of processing servers at each candidate node can reduce the energy penalty of the network at high demand levels. We also developed a heuristic for each scenario; EORIWG, EORIG and EORIGN for the scenario without geographical constraints, with geographical constraint and with geographical and link and node disjoint, respectively. The results show that the performance of the heuristic models is approaching the MILP models.

ACKNOWLEDGEMENT

The authors would like to acknowledge funding from the Engineering and Physical Sciences Research Council (EPSRC), INTERNET (EP/H040536/1), STAR (EP/K016873/1), Ministry of Education, Malaysia and Universiti Teknikal Malaysia Melaka (UTeM). All data is provided in the results section of this paper.

REFERENCES

[1] J. M. H. Elmirghani, T. Klein, K. Hinton, L. Nonde, A. Q. Lawey, T. E. H. El-Gorashi, M. O. I. Musa and X. Dong, "GreenTouch GreenMeter Core Network Energy-Efficiency Improvement Measures and Optimization," IEEE/OSA Journal of Optical Communication and Networking,, vol. 10, no. 2, pp. A250–A269, 2018.

[2] M. O. I. Musa, T. E. H. El-gorashi, and J. M. H. Elmirghani, "Bounds on GreenTouch GreenMeter Network Energy Efficiency," IEEE/OSA Journal of Lightwave Technology, vol. 36, no. 23, pp. 5395–5405, 2018.

[3] A. M. Al-Salim, A. Q. Lawey, T. E. H. El-Gorashi, and J. M. H. Elmirghani, "Energy Efficient Big Data Networks: Impact of Volume and Variety," IEEE Transaction on Network and Service Management, vol. 15, no. 1, pp. 458–474, 2018.

[4] M. Musa, T. Elgorashi, and J. Elmirghani, "Bounds for Energy Efficient Survivable IP Over WDM Networks with Network Coding," IEEE/OSA Journal of Optical Communications and Networking, vol. 10, no. 5, pp. 471–481, 2018.

[5] A. M. Al-salim, T. E. H. El-gorashi, A. Q. Lawey, and J. M. H. Elmirghani, "Greening big data networks: velocity impact," IET Optoelectronics, vol. 12, no. 3, pp. 126–135, 2018.

[6] A. N. Al-quzweeni, A. Q. Lawey, T. E. H. Elgorashi, and M. H. Jaafar, "Optimized Energy Aware 5G Network Function Virtualization," IEEE Access, vol. 7, pp. 44939–44958, 2019.

[7] M. S. Hadi, A. Q. Lawey, T. E. H. El-gorashi, and J. M. H. Elmirghani, "Big data analytics for wireless and wired network design: A survey," Computer Networks, vol. 132, pp. 180–199, 2018.

[8] X. Dong, A. Lawey, T. E. H. El-Gorashi, and J. M. H. Elmirghani, "Energy-efficient core networks," in 2012 16th International Conference on Optical Network Design and Modelling (ONDM), Colchester, pp. 1-9, 2012.

[9] X. Dong, T. El-Gorashi, and J. M. H. Elmirghani, "Green IP over WDM networks with data centers," IEEE/OSA Journal of Lightwave Technology, vol. 29, no. 12, pp. 1861–1880, 2011.

[10] H. M. Mohammad Ali, T. E. H. El-Gorashi, A. Q. Lawey, and J. M. H. Elmirghani, "Future Energy Efficient Data Centers With Disaggregated Servers," IEEE/OSA Journal of Lightwave Technology, vol. 35, no. 24, pp. 5361–5380, 2017.

[11] L. Nonde, T. E. H. El-gorashi, and J. M. H. Elmirghani, "Energy Efficient Virtual Network Embedding for Cloud Networks," IEEE/OSA Journal of Lightwave Technology, vol. 33, no. 9, pp. 1828–1849, 2015.

[12] A. Q. Lawey, T. E. H. El-Gorashi, and J. M. H. Elmirghani, "BitTorrent Content Distribution in Optical Networks," Journal Lightwave Technology, vol. 32, no. 21, pp. 4209–4225, 2014.

[13] N. I. Osman, T. El-Gorashi, L. Krug, and J. M. H. Elmirghani, "Energy-efficient future high-definition TV," Journal Lightwave Technology, vol. 32, no. 13, pp. 2364–2381, 2014.

[14] X. Dong, T. El-gorashi, and J. M. H. Elmirghani, "On the energy efficiency of physical topology design for IP over WDM networks," IEEE/OSA Journal of Lightwave Technology, vol. 30, no. 12, pp. 1931–1942, 2012.

[15] X. Dong, T. El-Gorashi, and J. M. H. Elmirghani, "IP Over WDM Networks Employing Renewable Energy Sources," IEEE/OSA Journal of Lightwave Technology, vol. 29, no. 1, pp. 3–14, 2011.

[16] M. Musa, T. Elgorashi, and J. Elmirghani, "Energy efficient survivable IP-Over-WDM networks with network coding," IEEE/OSA Journal Optical Communications and Networking, vol. 9, no. 3, pp. 207–217, 2017.

[17] Z. T. Al-Azez, A. Q. Lawey, T. E. H. El-Gorashi, and J. M. H. Elmirghani, "Energy Efficient IoT Virtualization Framework With Peer to Peer Networking and Processing," IEEE Access, vol. 7, pp. 50697–50709, 2019.

[18] B. G. Bathula, M. Alresheedi, and J. M. H. Elmirghani, "Energy Efficient Architectures for Optical Networks," Proceedings IEEE London Communications Symposium, London, pp. 5-8, 2009.

[19] B. G. Bathula and J. M. H. Elmirghani, "Energy efficient Optical Burst Switched (OBS) networks," IEEE Globecom Workshops, Honolulu, pp. 1-6, 2009.

[20] T. E. H. El-Gorashi, X. Dong, and J. M. H. Elmirghani, "Green optical orthogonal frequency-division multiplexing networks," in IET Optoelectronics, vol. 8, no. 3, pp. 137–148, 2014.

[21] I. S. M. Isa, T. E. H. El-Gorashi, M. O. I. Musa, and J. M. H. Elmirghani, "Energy Efficient Fog based Healthcare Monitoring Infrastructure," Arvix 2020.

[22] A. Q. Lawey, T. E. H. El-Gorashi, and J. M. H. Elmirghani, "Distributed energy efficient clouds over core networks," Journal of Lightwave Technology, vol. 32, no. 7, pp. 1261–1281, 2014.

[23] M. S. Hadi, A. Q. Lawey, T. E. H. el-gorashi, and J. M. H. Elmirghani, "Patient-Centric Cellular Networks Optimization using Big Data Analytics," IEEE Access, vol. 7, pp. 49279–49296, 2019.

[24] F. Jalali, K. Hinton, R. S. Ayre, T. Alpcan, and R. S. Tucker, "Fog Computing May Help to Save Energy in Cloud Computing," IEEE Journal on Selected Areas in Communication, vol. 34, no. 5, pp. 1728–1739, 2016.

[25] F. Jalali, A. Vishwanath, J. De Hoog, and F. Suits, "Interconnecting FOG computing and microgrids for greening IoT," 2016 IEEE Innovative Smart Grid Technologies - Asia (ISGT-Asia), Melbourne, VIC, pp. 693–698, 2016.

[26] I. S. M. Isa, M. O. I. Musa, T. E. H. El-gorashi, A. Q. Lawey, and J. M. H. Elmirghani, "Energy Efficiency of Fog Computing Health Monitoring Applications," in 2018 20th International Conference on Transparent Optical Networks (ICTON), Bucharest, pp. 1–5, 2018.

[27] C. Colman-meixner, C. Develder, M. Tornatore, and B. Mukherjee, "A Survey on Resiliency Techniques in Cloud Computing Infrastructures

and Applications," IEEE Communications Surveys & Tutorials, vol. 18, no. 3, pp. 2244–2281, 2016.

[28] R. S. Couto, S. Secci, M. E. M. Campista, and L. H. M. K. Costa, "Server placement with shared backups for disaster-resilient clouds," Computer Networks, vol. 93, pp. 423–434, 2015.

[29] C. Develder, J. Buysse, B. Dhoedt, and B. Jaumard, "Joint dimensioning of server and network infrastructure for resilient optical grids/clouds," in IEEE/ACM Transactions on Networking, vol. 22, no. 5, pp. 1591–1606, 2014.

[30] A. Modarresi and J. P. G. Sterbenz, "Toward Resilient Networks with Fog Computing," 2017 9th International Workshop on Resilient Networks Design and Modeling (RNDM), Alghero, pp. 1–7, 2017.

[31] I. S. M. Isa, M. O. I. Musa, T. E. H. El-Gorashi, and J. M. H. Elmirghani, "Energy efficient and resilient infrastructure for fog computing health monitoring applications," in 2019 21st International Conference on Transparent Optical Networks (ICTON), Angers, France, pp. 1-5, 2019.

[32] Blueraq Networks Ltd, "Cloud Backup: UK Based, Automatic & Secure," BACKUP VAULT, 2019. [Online]. Available: https://www.backupvault.co.uk/ BackupVault.

[33] Public Health England, "Public Health Profiles," Public Health England, 2015. [Online]. Available: http://healthierlives.phe.org.uk/. [Accessed: 20-Jun-2015].

[34] N. Hannent, "Ofcom UK Mobile Sitefinder," Ofcom UK Mobile. [Online].Available:https://fusiontables.google.com/DataSource?docid =1N4nf1AmXFDk-Ibh9Y54jh2FwyudbX3O8-aVlzwZJ#rows:id=1.

[35] "FTTC Exchanges," Sam Knows Ltd 2019, 2016. [Online]. Available: https://availability.samknows.com/broadband/exchanges/bt/fttc.

[36] "Cardiovascular Disease Statistic 2017", British Heart Foundation, 2017.

[37] Cisco, "The Zettabyte Era : Trends and Analysis," 2017.

[38] R. Prieto, "Cisco Visual Networking Index Predicts Near-Tripling of IP Traffic by 2020," 2016.

[39] G. B. Moody and R. G. Mark, "The impact of the MIT-BIH Arrhythmia Database," in IEEE Engineering in Medicine and Biology Magazine, vol. 20, no. 3, pp. 45-50, 2001.

[40] A. L. Goldberger, L A Amaral, L Glass, J M. Haundorff, P C Ivanov, R G Mark, J E Mietus, G B Moody, C K Peng and H E Stanley, "PhysioBank, PhysioToolkit, and PhysioNet Components of a New Research Resource for Complex Physiologic Signals," Circulation, vol. 101, no. 23, pp. 215-220, 2000.

[41] Stone, "STONEPC LITE," Stone Group, 2017. [Online]. Available: https://www.stonegroup.co.uk/hardware/desktops/lite/.

[42] Alcatel-Lucent, "Alcatel-Lucent 7368 ISAM ONT G-240G-A," 2014.

[43] Apple Support, "iMac power consumption and thermal output," 2017. [Online]. Available: https://support.apple.com/en-gb/HT201918.

[44] V. Valancius, N. Laoutaris, L. Massoulié, C. Diot, and P. Rodriguez, "Greening the internet with nano data centers," in 5th international conference on Emerging networking experiments and technologies CoNEXT 09, pp. 37-48, 2009.

[45] R. W. A. A. and R. S. T. A. Vishwanath, F. Jalali, K. Hinton, T. Alpcan, "Energy consumption comparison of interactive cloud-based and local applications," IEEE Journal on Selected Areas in Communications, vol. 33, no. 4, pp. 616–626, 2015.

[46] P. Mahadevan, P. Sharma, S. Banerjee, and P. Ranganathan, "A power benchmarking framework for network devices," International Conference on Research in Networking, pp. 795–808, 2009.

[47] Nokia, "LTE-M – Optimizing LTE for the Internet of Things White Paper," 2015.

[48] J. Baliga, R. W. A. Ayre, K. Hinton, and R. S. Tucker, "Green Cloud Computing: Balancing Energy in Processing, Storage and Transport," in Proceedings of the IEEE, vol. 99, no. 1, pp. 149–167, 2011.

[49] C. Gray, R. Ayre, K. Hinton, and R. S. Tucker, "Power consumption of IoT access network technologies," in IEEE International Conference on Communication Workshop (ICCW), pp. 2818–2823, 2015.

[50] Eltex, "GPON Optical Line Terminal Data Sheet," 2015.

[51] G. Auer, O. Blume, V. Giannini, I. Godor, and M. A. Imran, "Energy efficiency analysis of the reference systems, areas of improvements and target breakdown," EARTH Project Report Deliverable D2.3, pp. 1–68, 2012.